\documentclass[%
 reprint,
 amsmath,amssymb,
 aps,
]{revtex4-2}

\usepackage[colorlinks=true,linkcolor=blue,citecolor=blue,urlcolor=blue]{hyperref}

\usepackage{graphicx}
\usepackage{dcolumn}
\usepackage{bm}
\usepackage[mathlines]{lineno}

\usepackage{soul}
\usepackage{yfonts}
\usepackage{amsmath}
\usepackage{xcolor}
\usepackage{float}
\usepackage{amsfonts}
\usepackage{amsthm}

\begin{document}

\preprint{APS/123-QED}

\title{Super-resolution and super-sensitivity of quantum LiDAR with multi-photonic state and binary outcome photon counting measurement}

\author{ Priyanka~Sharma$^{1}$}
\author{ Manoj~K~Mishra$^{2}$}
\author{ Devendra~Kumar~Mishra$^1$}
\email{Corresponding author: kndmishra@gmail.com}
\affiliation{$^1$Deptartment of Physics, Banaras Hindu University, Varanasi-221005, India}
\affiliation{$^2$Space Applications Centre, Indian Space Research Organization (ISRO), Ahmedabad, Gujarat, India}

\vspace{10pt}

\date{\today}

\begin{abstract}
Here we are investigating the enhancement in resolution and phase sensitivity of a Mach Zehnder interferometer (MZI) based quantum LiDAR. We are using multi-photonic state (MPS), superposition of four coherent states \cite{mishra2021ququats}, as the input state and binary outcome parity photon counting measurement and binary outcome zero-nonzero photon counting measurement as the measurement schemes. We have thoroughly investigated the results in lossless as well as lossy cases. We found enhancement in resolution and phase sensitivity in comparison to the coherent state and even coherent superposition state (ECSS) based quantum LiDAR. Our analysis shows that MPS may be an alternative nonclassical resource in the field of quantum imaging and quantum sensing technologies, like in quantum LiDAR.
\end{abstract}

\maketitle

 \section{Introduction}\label{section1}
RADAR (radio detection and ranging) \cite{ref1} and its optical counterpart, LiDAR (light detection and ranging) \cite{Collis:70}, are well-established and broadly used technologies. In RADAR, radio frequencies are used for long-range detection \cite{kitchen1993weather}, while in LiDAR, visible to near-infrared frequencies are used for detection and ranging at a shorter distance \cite{lin2019evaluation, slepyan2021quantum}.

LiDAR, a combination of laser technology and modern photoelectric detection technology, has advantages over traditional radar such as high resolution \cite{wallace2012assessing, blakey2022quantum, liu2023compact}, strong anti-interference ability \cite{tsai2020anti}, and good low-altitude detection performance \cite{laurenzis2017multi}. However, with continuous application in aerospace, military, artificial intelligence and other fields, LiDAR has reached the limits of classical physics \cite{kim2021nanophotonics}. With the emergence of quantum metrology \cite{1976Helstrom, giovannetti2006quantum}, the quantum version of LiDAR has been proposed to achieve better precision and resolution in comparison to classical LiDAR \cite{dowling2009quantum, slepyan2021quantum}.

In quantum LiDAR, resolution and phase sensitivity are the important parameters to determine the range and image of the target \cite{dowling2009quantum, doi:10.1080/00107510802091298}. Since the fringe resolution is determined by the `full width at half maximum' ($\text{FWHM}$) of the intensity signal i.e, $\Delta x \propto \text{FWHM}$. A laser light (coherent state) assisted conventional MZI gives the fringe resolution ($\Delta x$) of $\lambda/2$ in intensity measurement, where $\lambda$ represents the wavelength of light.  This is the maximum resolution achieved in the classical regime termed as the Rayleigh resolution limit \cite{born2013principles, boto2000quantum}. Beating the Rayleigh limit is termed as the super-resolution and one can achieve this by using the coherent state with quantum detection schemes \cite{mitchell2004super, PhysRevLett.98.223601}. For example, Gao \textit{et al.} \cite{Gao:10} presented a super-resolving quantum LiDAR method with coherent states (CS) and photon-number-resolving detectors and Distante \textit{et al.} \cite{PhysRevLett.111.033603} established the super-resolution with $x \propto \lambda/2N$. Next to the resolution, the coherent state-assisted MZI also gives the ultimate phase sensitivity achieved by the quantum theory. This limit is known as shot-noise limit (SNL) and is given by $\Delta\phi_{SNL}={1}/{\sqrt{\Bar{n}}}$, where $\Bar{n}$ is the mean photon number of the coherent state. 

Rayleigh limit and SNL can be surpassed by utilizing the non-classical resources \cite{dowling2009quantum, PhysRevD.23.1693, xiao1987precision, shukla2024quantum, shukla2023improvement, dowling2008quantum, Shukla:22, boto2000quantum, giovannetti2011advances, Shukla:21} and quantum detection techniques \cite{Gao:10, PhysRevLett.111.033603, mitchell2004super, PhysRevLett.98.223601}.  Dowling's group \cite{dowling2009quantum} showed that by using N00N state, \cite{boto2000quantum} surpasses the Rayleigh limit and the SNL. However, in the presence of loss and noise, the N00N states are highly fragile making it difficult to achieve the super-phase sensitivity and super-resolution \cite{PhysRevA.78.063828, PhysRevA.75.053805}. Reichert \textit{et al.} \cite{reichert2022quantum} showed that using squeezed and frequency entangled signal and idler beams precision enhancement in the estimation of the velocity of the object can be achieved. In article \cite{WANG20163717}, Wang \textit{et al.} showed that with even coherent superposition states (ECSS), defined as $(|i\alpha\rangle + |-i\alpha\rangle)$ \cite{DODONOV1974597}, they got the improvement in resolution and sensitivity compared to the CS as an input by using binary-outcome photon counting measurements even in small photon loss. Apart from resolution and phase sensitivity, the superposition of coherent states plays a vital role in quantum technologies and quantum teleportation \cite{mishra2010teleportation, prakash2012teleportation, mishra2016quantum, maurya2016two, mishra2011two, mishra2013bipartite}.

In this paper, we are considering the more general superposition of coherent states termed as `multi-photonic state (MPS)' proposed by Mishra \textit{et al.} \cite{mishra2021ququats} for investigating the possibility of super-sensitivity and super-resolution in quantum LiDAR. MPS is defined as\cite{mishra2021ququats}
\begin{equation}
  |\Psi_j\rangle = N_j\sum^{3}_{m=0}(-i)^{jm}|(i)^m\alpha\rangle,~~~j=0,1,2,3\label{eq1}
\end{equation} 

and the normalization constant is
\begin{equation}
 \begin{split}
 N_{0,2}= \left(4(1+r^2\pm2r \cos{|\alpha|^2})\right)^{\frac{-1}{2}},\\
 N_{1,3}= \left(4(1-r^2\pm2r \sin{|\alpha|^2})\right)^{\frac{-1}{2}}
 \end{split}
\end{equation}
where $r=exp(-|\alpha|^2)$ and $|\alpha\rangle = exp(-|\alpha|^2/2)\sum^{\infty}_{l=0}(\alpha^{l}/\sqrt{l!})|l\rangle$ is the usual Glauber coherent state of radiation with coherent amplitute $|\alpha|$. 
MPS contains $(4n+j)$ photons, where $j$ ranges from $0, 1, 2, 3$ and $n$ ranges from $0, 1, ...,\infty$. Different nonclassical properties of MPS applicable in quantum technology have been studied and an experimental scheme for the generation of MPS has been proposed by Mishra \textit{et al.} in \cite{mishra2021ququats}. In this scheme, the even superposition of coherent states serves as the inputs of the MZI, having a phase shifter in one of the arms, and on the output we get the MPS. Since generating even superposition coherent states is feasible with current technology, producing MPS is also achievable.

The motivation for considering MPS for investigation of the super-resolution and super-phase sensitivity of a quantum LiDAR is comes from the result achieved by ECSS. ECSS, is the superposition of two coherent states while, in MPS, we take the superposition of four coherent states. Also, we can see that ECSS is a special case of MPS. Therefore, analysis with MPS will give us a vast canvas to look into the possibility of finding the super-resolution and super-phase sensitivity in the field of quantum LiDAR. For detection purposes, we choose two famous, binary outcome parity photon counting measurement and binary outcome zero non-zero photon counting measurement, schemes. In the rest of the manuscript, we call them parity detection and Z-detection schemes, respectively.

The paper is organized as follows. In section \ref{section 2}, we discuss the special cases of MPS along with their Wigner function and Section \ref{section 3}, explains about the MZI as a quantum LiDAR. We also discuss the parameter estimation and the detection schemes. Section \ref{section 4} contains the observation with MPS and vacuum state as the inputs of the MZI and Section \ref{section 5} contains the observation with MPS and coherent state. In Section \ref{section 6}, we conclude our results.

\section{Wigner function of MPS }\label{section 2}
In this section, we investigate the nonclassical witness of the MPS with the help of its Wigner function distribution.

 MPS is defined in Eq. \eqref{eq1} and can be rewrite as
\begin{equation}
|\Psi_j\rangle= N_j(A|\alpha\rangle+B|i\alpha\rangle+C|-\alpha\rangle+D|-i\alpha\rangle),\label{eq2}
\end{equation}
where $A=1,~B=(-i)^j,~C=(-i)^{2j},~D=(-i)^{3j}$. 

From Eq. \eqref{eq2}, using the relation $\langle\alpha|\beta\rangle = exp(\frac{-|\alpha|^2-|\beta|^2+2\alpha^*\beta}{2})$, normalisation constant can be written as
\begin{equation}
 \begin{split}
 |N_j|= \left(X +2Ye^{-|\alpha|^2}\cos(j\pi)\right. \\
 \left.+2Ve^{-|\alpha|^2}\cos\left(|\alpha|^2-j\frac{\pi}{2}\right)\right)^{\frac{-1}{2}},
 \end{split}\label{17a}
\end{equation}
where,

   $ X=(|A|^2+|B|^2+|C|^2+|D|^2)$,
   $Y=(|A||C|+|B||D|),~V=(|A|+|C|)(|B|+|D|)$.\label{eq5}
In Eq. \eqref{eq2}, for four different values of $j=0,~1,~2,$ and $3$, we get the four different multi-photonic states and we refer them as $\text{MPS}_{j=0}$, $\text{MPS}_{j=1}$, $\text{MPS}_{j=2}$ and $\text{MPS}_{j=3}$, respectively, in our further discussion.

Wigner function \cite{wigner1932quantum, hillery1984distribution} is a phase-space distribution function for a quantum system. The Wigner function has turned out to be remarkably useful to characterize and visualize the nonclassicality of a quantum state. The value of the Wigner function lies between $-1$ and $1$, i.e., $-1\leq W\leq 1$. For a non-classical state, the Wigner function is always $<0$.

The Wigner function of a state with density operator $\hat{\rho}$, can be defined as
 \begin{equation}
 W(\lambda)= \frac{1}{\pi^2}\int d^2\eta \text{Tr}[\hat{\rho} \hat{D}(\eta)]e^{-(\eta\lambda^{*}-\eta^{*}\lambda)}.\label{wf7}
 \end{equation}
Where $\lambda$ and $\eta$ are the complex variables and $\hat{D}(\eta)$ is the displacement operator. The density operator for a pure state, $|\psi\rangle$, can be written as $\hat{\rho}= |\psi\rangle\langle\psi|$. Since, the displacement operator is defined as $\hat{D}(\eta) = exp{(\eta\hat{a}^\dagger-\eta^*\hat{a})}$, therefore,  
\begin{equation}
   \text{Tr}[\hat{\rho}\hat{D}(\eta)] =  \langle \psi|\exp{(\eta\hat{a}^\dagger-\eta^*\hat{a})}|\psi\rangle.\label{eq6wig}
\end{equation}
In order to calculate the Wigner function of MPS we calculate the trace (Eq. \eqref{eq6wig}) with respect to $|\Psi_j\rangle$ (Eq. \eqref{eq2}) and put them in Eq. \eqref{wf7}. For detailed calculation go through Appendix \ref{appendix wigner}. After a straightforward calculation, we get the Wigner for the MPS, which is written as
\begin{widetext}
    \begin{equation}
 \begin{split}
      W_j(\lambda)= \frac{2N_j}{\pi}\left(e^s(|A|^2e^{4p_1}+|B|^2e^{4q_1}+|C|^2e^{-4p_1}+|D|^2e^{-4q_1})+2e^{(u+v)}(|A||B|\cos(u+u')\right.\\
      +|A||D|\cos(-u+u'))+2e^{-2|\lambda|^2}(|A||C|\cos(4q_1-j\pi )+|B||D|\cos(4p_1-j\pi))\\
      \left.+2|C||D|e^{(u-4p_1+v)}\cos(u-4p_1-u')+2|B||C|e^{(-u+v)}\cos(u+u')\right),
 \end{split}\label{eq9}
\end{equation}
\end{widetext}
where,
 \begin{equation}
     \begin{split}
u=2(p_1-q_1),~v=-2|\lambda|^2-|\alpha|^2,~u'=|\alpha|^2+\frac{j\pi}{2},\\
\alpha=x_1+ix_2,~\lambda=y_1+iy_2,~
p_1 =x_1y_1+x_2y_2,\\
q_1 =x_1y_2-x_2y_1,~s=-2(|\alpha|^2+|\lambda|^2).
     \end{split}
 \end{equation}
Where $x_1,~y_1$ and $x_2,~y_2$ are the real and imaginary part of the $\alpha,~\lambda$, respectively. Since, Eq. \eqref{eq9} is the Wigner function of MPS, so, we can write the Wigner function for $\text{MPS}_{j=0}$, $\text{MPS}_{j=1}$, $\text{MPS}_{j=2}$ and $\text{MPS}_{j=3}$, by putting appropriate values of the coefficients $A,~B,~C,~D$ and $j$. In order to compare the results, we also plot the Wigner functions of the CS and ECSS states.

\begin{figure*}
\includegraphics[width=18 cm, height=7.5cm]{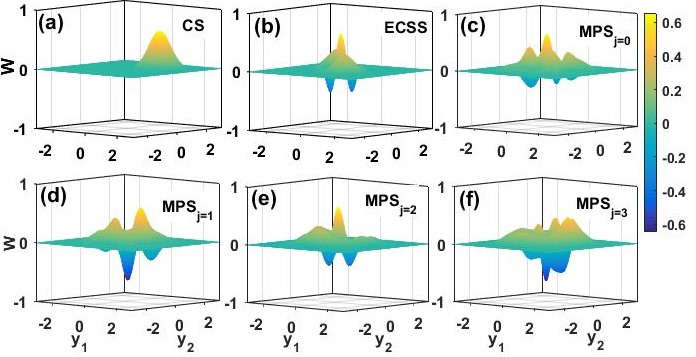}
\caption{\label{fig:2(C)} Plots show the Wigner function distribution for all six states. The colour bar shows the corresponding values of the Wigner function. The negative values show the condition on $y_1$ and $y_2$ for a state to become a nonclassical state. Here, we consider $x_1=x_2=1$.}
\end{figure*}

We plot the Wigner function for all the states with respect to $y_1$ and $y_2$, by considering, $x_1=x_2=1$, as shown in Fig. \ref{fig:2(C)}.
From Fig. \ref{fig:2(C)}, we can see that for CS, the Wigner function always $\geq0$, which shows that CS is a classical state. On the other hand, the Wigner function of the other five states (ECSS, $\text{MPS}_{j=0},~\text{MPS}_{j=1},~\text{MPS}_{j=2}$ and $\text{MPS}_{j=3}$) showing negative values. Here, $\text{MPS}_{j=1}$ and $\text{MPS}_{j=3}$ shows high negativity in comparison to the other states. Hence, the negativity of the Wigner function provides the signature of nonclassicality of $\text{MPS}_{j=0,1,2,3}$. Also, we can get the signature of nonclassicality for some other values of $x_1$ and $x_2$.

\section{Mach-Zehnder Interferometer as a Quantum LiDAR}\label{section 3}
\subsection{Basic working and state transformation}
The basic working principle of a quantum LiDAR depends on the principle of MZI. The MZI is widely used in quantum sensing and imaging technologies \cite{dowling2009quantum}. Therefore, here we briefly discuss the working principle of MZI and state transformation.

\begin{figure}
\includegraphics[width=8.5cm, height=5.5cm]{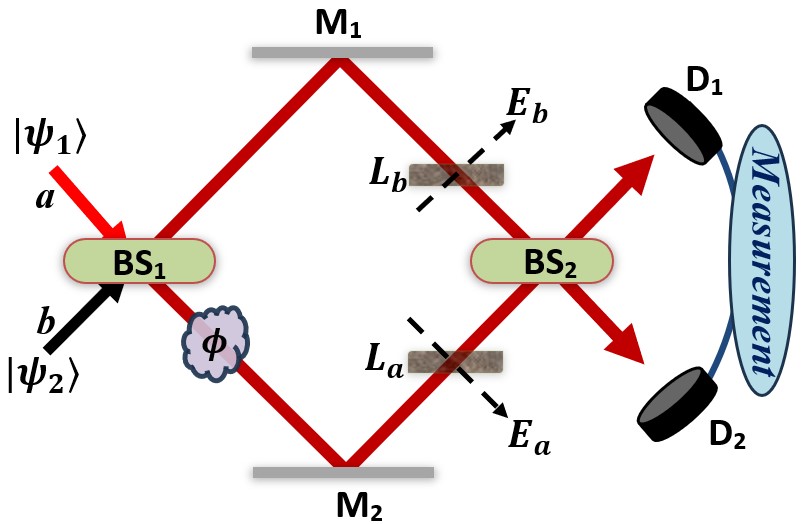}
\caption{\label{fig:1} MZI has two input (modes $a,~b$) and two output ports including two 50:50 beam splitters ($BS_1$ and $BS_2$), two mirrors ($M_1$ and $M_2$), a phase shifter $\phi$, two detectors $(D_1 \& D_2)$. The two fictitious beam splitters  ($L_a$ and $L_b$) whose corresponding modes are $E_a$ and $E_b$, respectively, mimic the photon loss inside the two arms of MZI.}
\end{figure}

Working of a typical MZI setup (shown in Fig. \ref{fig:1}) can be understood in three steps: probe preparation, probe evolution, and measurement. In probe preparation, two input states are combined by a beam splitter where operator transformation follows the SU(2) algebra, the input-output relation can be written as
\begin{equation}
    \begin{pmatrix}
    \hat{a}_{out}\\
    \hat{b}_{out}
\end{pmatrix}=\textbf{B}
\begin{pmatrix}
    \hat{b}_{in}\\
    \hat{a}_{in}
\end{pmatrix},
\label{eq:3}
\end{equation}
where $\hat{a}_{in}, \hat{b}_{in}$ ($\hat{a}_{out}, \hat{b}_{out}$) are the input (output) annihilation operators of the beam splitter. \textbf{B} is the transformation matrix of the beam splitter and is defined as $\textbf{B}=\begin{pmatrix}
    i\mathcal{\sigma}&\mathcal{\tau}\\
    \mathcal{\tau}&i\mathcal{\sigma}
\end{pmatrix}$, where $\mathcal{\sigma}$ and $\mathcal{\tau}$ are the reflectance and transmittance of the beam splitter and follow the relation $\mathcal{\sigma}^2 + \mathcal{\tau}^2=1$. For a 50:50 beam splitter, $\mathcal{\sigma} = \mathcal{\tau} ={1}/{\sqrt{2}}$, the phase change being $\pi/2$ (0) in reflection (transmission). So, at the outputs of $BS_1$, we get the superposition of the input states and we call one as reference light and the second as probe light. Probe light encodes the information of our interest, in the case of quantum LiDAR, this information is about the distant object. The phase shift $\phi$ experienced in the path of the interferometer can be expressed in terms of the unitary evolution operator $\hat{U}(\phi) = e^{i\phi\hat{a}^\dagger\hat{a}}$ with $\phi= 2kf$, where $f$ is the target distance and $k = \frac{2\pi}{\lambda}$ the wave number ($\lambda$ is the wavelength of the light). During the evolution process, the probe signal interacts with the environment which causes photon loss in the signal. This photon loss occurs mainly because of the scattering and reflection of light from the different types of objects present in the path. So, in order to consider the photon loss inside the LiDAR, we consider two fictitious beam splitters ($L_a$ and $L_b$) in each arm having transmittivity $t$ and reflectivity $r$ for mimicking the photon loss\cite{loudon2000quantum, Agarwal_2012}. After the evolution of the probe, we combine the probe light on the second beam splitter $BS_2$ with the reference light. In the last step, we carry out the measurement on the output of the $BS_2$ by detector$D_1$ and $D_2$.

So, initially, we consider the state of the system as
\begin{equation}
    |\psi\rangle_{in}
=|\hat{a},\hat{b},0,0\rangle_{a,b,E_a,E_b}.
\end{equation}
Here, $\hat{a}$ and $\hat{b}$ are the annihilation operators of the inputs and $a$, $b$ and $E_a$, $E_b$ are the modes of input states and photon loss (as shown in Fig. \ref{fig:1}).
Here, the transformation matrices associated with 50:50 beam splitters $BS_1$ and $BS_2$ are $\hat{B_1}$ and $\hat{B_2}$, respectively, such that $\hat{B_1} = \hat{B_2} = \frac{1}{\sqrt{2}}\begin{pmatrix}
    i&1\\
    1&i
\end{pmatrix}$. The transformation matrices associated with fictitious beam splitters $L_a$ and $L_b$ are $\hat{B}_{l_a}$ and
 $\hat{B}_{l_b}$ such that  $\hat{B}_{l_a} =\hat{B}_{l_b} = \begin{pmatrix}
    ir&t\\
    t&ir
\end{pmatrix}$. We can write the phase shift operator in the matrix as
$\hat{U}(\phi) = \begin{pmatrix}
    e^{i\phi}&0\\
    0&1
\end{pmatrix}$. Therefore after the all transformations, we can write the output state as
\begin{equation}
\begin{split}
|\psi\rangle_{out} = \left|\psi_{1_{out}}, \psi_{2_{out}},\frac{ir(\hat{a}+i\hat{b})e^{i\phi}}{\sqrt{2}},\frac{ir(\hat{b}+i\hat{a})}{\sqrt{2}}\right\rangle_{a,b,E_a,E_b},
\end{split}\label{3}
\end{equation}
where, $\psi_{1_{out}}=\frac{1}{\sqrt{2}}\left(\frac{t(\hat{a}+i\hat{b})e^{i\phi}}{\sqrt{2}}+\frac{it(\hat{b}+i\hat{a})}{\sqrt{2}}\right)$ and $\psi_{2_{out}}=\frac{1}{\sqrt{2}}\left(\frac{it(\hat{a}+i\hat{b})e^{i\phi}}{\sqrt{2}}+\frac{t(\hat{b}+i\hat{a})}{\sqrt{2}}\right)$.

\subsection{Parameter estimation}
In order to perform the measurement on the output of the interferometer, we must have an observable  corresponding to the detection scheme. Let us consider $\hat{X}(\phi)$ as an observable that contains the information of the unknown phase $\phi$ to be estimated. Here in our case $\hat{\Pi}$ and $\hat{Z}$ are the observables for parity and $Z$- detection schemes respectively.

A small change of $\delta\phi$ in $\phi$ rises a change in $\hat{X}(\phi)$ to $\hat{X}(\phi+\delta \phi)$. From Taylor's expansion the expectation value of $\hat{X}(\phi+\delta \phi)$ can be written as 
\begin{equation}
\langle\hat{X}(\phi+\delta \phi)\rangle \approx \langle\hat{X}(\phi)\rangle + \frac{\partial\langle\hat{X}(\phi)\rangle}{\partial\phi}\delta\phi.
\end{equation}
The difference $\langle\hat{X}(\phi+\delta \phi)\rangle - \langle\hat{X}(\phi)\rangle$ is detected in the experiment only if 
\begin{equation}
\langle\hat{X}(\phi+\delta \phi)\rangle - \langle\hat{X}(\phi)\rangle \geq \Delta \hat{X}(\phi)\label{5p},
\end{equation}
where, $\Delta\hat{X}(\phi) = \sqrt{\langle\hat{X}^2\rangle -\langle\hat{X}\rangle^2}$ is the standard deviation of $\hat{X}(\phi)$. The value of $\delta\phi$ for which inequality given in Eq. \eqref{5p} saturates is called the phase sensitivity $\Delta\phi$ and can be written as \cite{demkowicz2015quantum, paris2009quantum}
\begin{equation}
\Delta\phi = \frac{\Delta\hat{X}}{\left|\frac{\partial\langle\hat{X}(\phi)\rangle}{\partial\phi}\right|}\label{6p}.
\end{equation}
This equation will be vital in the following sections for the sensitivity of the MZI.

\subsection{Detection schemes}
Here we will briefly introduce the detection schemes considered in our analysis.
\subsubsection{\textbf{Parity detection}}\label{ppc}
The operator for the parity measurement is defined as $\hat{\Pi} = (-1)^{\hat{n}}= e^{i\pi\hat{n}}$, where $\hat{n}=\hat{a}^\dagger\hat{a}$ is the number operator  \cite{royer1977wigner, cahill1969density, gerry2007parity, PhysRevA.54.R4649}. So, on the basis of photon counts on the output port, we have $\hat{\Pi}=+1$ or $\hat{\Pi}=-1$ for an even number of photons or an odd number of photons, respectively. So, the expectation value of parity operator $\hat{\Pi}$ is obtained by
\begin{equation}
    \langle\hat{\Pi}\rangle = \sum_{n=0}^{\infty} (-1)^n P(n),
\end{equation}
where $P(n)$ is the probability for $n$ photons in the output state. If the probability of getting an even number of photons, is $P(+)$ and the odd number of photons is $P(-)$ with  $P(+)+P(-)= 1$, then the expectation value of parity operator in terms of even and odd probabilities can be written as
\begin{equation}
\langle\hat{\Pi}\rangle = P(+)-P(-).\label{11pi}
\end{equation}
 
Using the error propagation formula as written in Eq. \eqref{6p}, phase sensitivity with the parity measurement  can be written as 
\begin{equation}
 \Delta\phi_{\Pi} = \frac{\sqrt{1-\langle \hat{\Pi}\rangle^2}}{\left|\frac{\partial\langle\hat{\Pi}\rangle}{\partial\phi}\right|}.\label{phi}
 \end{equation}
Here,  $\left|\frac{\partial\langle\hat{\Pi}\rangle}{\partial\phi}\right|$ is the variation in $\langle\hat{\Pi}\rangle$ with phase $\phi$ and we used the property $\hat{\Pi}^2=1$.
 
\subsubsection{\textbf{Z-detection}}\label{ZD}
In this measurement scheme, we explore the data obtained by a standard single-photon detector at the output port $a$ that cannot differentiate between different photon numbers \cite{feng2014quantum, cohen2014super}. It has only two outcomes, no photon ($n=0$) and any number of photons ($n\neq0$) with the associated probabilities
$P(n=0)$ and $P(n\neq0)=1-P(0)$, respectively. For a better approximation, we can take the difference between these two probabilities, i.e., $P(0)-P(n\neq0)$ \cite{cohen2013experimental}. This gives us $P(0)-P(n\neq0)=2P(0)-1$, so considering only $P(0)$ is much simpler to evaluate. Therefore, the observable for zero photon detection is the zero photon projection $\hat{Z}=|0\rangle_{a~a}\langle0|$, we call it Z-detection and its expectation value will be
\begin{equation}
     \langle \hat{Z}\rangle=P(0).\label{14Z}
\end{equation} 
So, the phase sensitivity in the case of Z-detection can be written as,
\begin{equation}
 \Delta\phi_Z = \frac{\sqrt{\langle \hat{Z}\rangle-\langle \hat{Z}\rangle^2}}{\left|\frac{\partial\langle \hat{Z} \rangle}{\partial\phi}\right|}.\label{15zd}
\end{equation}
Here, $\left|\frac{\partial\langle \hat{Z} \rangle}{\partial\phi}\right|$ is the variation in $\langle\hat{Z}\rangle$ with phase $\phi$ and we have used the property $\hat{Z}^2=\hat{Z}$.
 
We will use these two detection schemes for the super-resolution and super-sensitivity of quantum LiDAR.

\section{Observation with MPS and vacuum state as inputs}\label{section 4}
Here, we are discussing the resolution and phase sensitivity of the MZI by using MPS ($|\Psi_j\rangle$) and vacuum ($|0\rangle$) (as shown in Fig. \ref{fig:1}). Our initial state of the system can be written as 
\begin{equation}
|\psi\rangle_{in}=|\Psi_j,0,0,0\rangle_{a,b,E_a,E_b}.
\end{equation}
Here  $|\Psi_j\rangle$ is the MPS at input mode $a$ as given in Eq. \eqref{eq2} whereas $b,~E_a$ and $E_b$ are the respective modes of the system in vacuum state as shown in Fig. \ref{fig:1}. Then, using Eq. \eqref{3}, the final state of the system at the output of MZI becomes
\begin{equation}
\begin{split}
     |\psi\rangle_{out}= N_j\left(A|K\rangle
     +B|L\rangle +C|M\rangle
+D|N\rangle\right)_{a,b,E_a,E_b}.\label{20out}
\end{split}
\end{equation}
Here $|K\rangle,~|L\rangle,~|M\rangle,~|N\rangle$ are given in Eq. \eqref{eqA3}. Eq. \eqref{20out}, contains the information about the unknown phase change during the probe evolution inside the interferometer and this will be used to calculate the resolution and phase sensitivity of the MZI in this case. For this purpose, we consider two measurement schemes defined in terms of two observables $\hat{\Pi}$ and $\hat{Z}$. The expectation values of $\hat{\Pi}$ and $\hat{Z}$ are calculated in terms of the photon number probability as already discussed in Sections \ref{ppc} \& \ref{ZD}. To do this, the density operator for the final state of the system can be written as,
\begin{equation}
    \hat{\rho}_{out}= |\psi\rangle_{out~out}\langle\psi|,\label{22abc}
\end{equation}
where $|\psi\rangle_{out}$ is given in Eq. \eqref{20out}. For detailed expression go through the Appendix \ref{appendix a}. So, the probability of detecting $n$ photons at the output port $a$ and $m$ photons at the output port $b$ is given by 
\begin{equation}
 P(n,m)=\langle n,m|\hat{\rho}_{out}|n,m\rangle \label{eq6}.
\end{equation}
In order to find the probability of getting $n$ photon number at port $a$, we sum over $m$ and can be written as 
\begin{equation}
P(n)=\sum_{m}P(n,m).\label{p9}
\end{equation}
Therefore, from Eqs. (\ref{eq6}) and (\ref{p9}), the probability of getting $n$ photon numbers at output port $a$ is 
\begin{equation}
 \begin{split}
P(n) =|N_j|^2e^{-|\alpha|^2} \left(Xe^{-p+|\alpha|^2}\frac{(p)^n}{n!}\right.\\
+V\left(e^{-iq}\frac{(-ip)^n}{n!}+c.c.\right)\left.+Y\left(e^{q'}\frac{(-p)^n}{n!}+c.c.\right)\right).
 \end{split}  \label{eq p18}
\end{equation}
Where
\begin{equation}
    \begin{split}
        q=\left(|\alpha|^2x-j\frac{\pi}{2}\right),q'=(-|\alpha|^2x+ij\pi),\\
        x=|t|^2cos^2\left(\frac{\phi}{2}\right)+|r|^2,~p= |\alpha|^2|t|^2sin^2\left(\frac{\phi}{2}\right),\label{eq29}
    \end{split}
\end{equation}
and $V,~X$ and $Y$ are given in Section \eqref{eq5}.
\subsection{Parity detection}
In parity detection, the parity operator, $\hat{\Pi}$, divides the photon counting data $\{n,m\}$ into binary outcomes $\pm$, according to the even or odd number of photons at the output port. So, from Eq. \eqref{eq p18} the probability of getting an even or odd number of photons is
 \begin{equation}
 \begin{split}
P(\pm)=|N_j|^2 e^{-|\alpha|^2}\left(\frac{1}{2}Xe^{|\alpha|^2}(1\pm e^{-2p})+V(cos(q+p)\right.\\
\pm cos(q-p))\left.+Ye^{-|\alpha|^2}\left((1\pm e^{2p})cos(j\pi)\right)\right).
 \end{split}   
\end{equation}
For detail expression go through Appendix \ref{appendix a2}. Here, $p,~q,~q'$ are defined in Eq. \eqref{eq29}.
From the definition of average of parity, given in Eq. \eqref{11pi}, the parity measurement for the system  can be written as 
 \begin{equation}
 \begin{split}
\langle\hat{\Pi}\rangle =|N_j|^2 e^{-|\alpha|^2}(X e^{-(2p-|\alpha|^2)}\\+2Vcos(q-p)+Ye^{(2p-|\alpha|^2)}cos(j\pi)).\label{32pi}
 \end{split}   
\end{equation}
\begin{figure*}
\includegraphics[width=18.5cm, height=8.5cm]{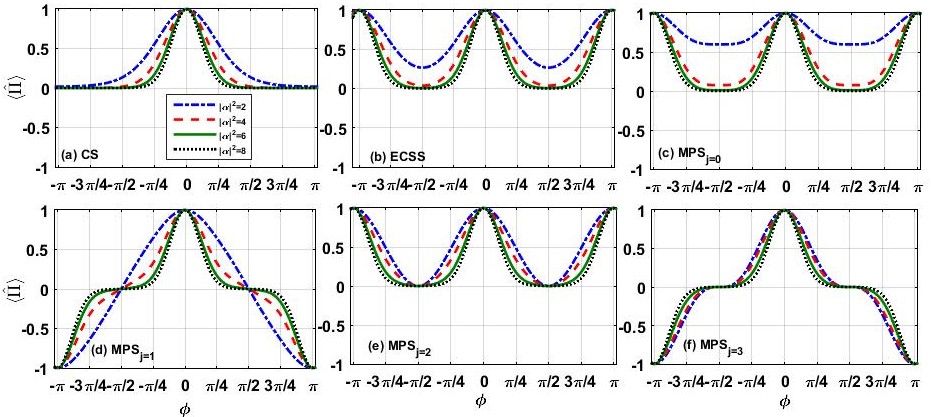}
\caption{\label{fig:5(a)} This figure shows the variation of $\langle\hat{\Pi}\rangle$ with respect to $\phi$ for different values of $|\alpha|^2$ for all six chosen states. We can see that the FWHM of the different states decreases with increasing $|\alpha|^2$.}
\end{figure*}
This result in Eq. \eqref{32pi} will be used to calculate the resolution and phase sensitivity.

In direct detection, the resolution of the system depends on the `full width at half maximum (FWHM)' of the intensity signal, i.e., $\Delta x \propto \text{FWHM}$. So, here, we follow the same approach for discussing the resolution of the MZI.  To measure the resolution, with the help of the oscilloscopes, intensity patterns are noted down and using FWHM, we estimate the resolution.
\begin{figure}
\includegraphics[width = 8.5cm, height = 10.5cm]{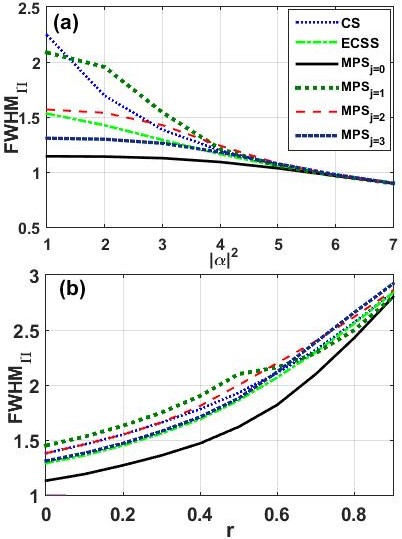}\caption{\label{fig_6}(a) Plot shows the variation of FWHM with $|\alpha|^2$ for the different states and (b) shows the loss tolerance with mean photon number $N=3$ in parity detection.}
\end{figure}

\begin{figure}[H]
\includegraphics[width=8.5cm, height= 6.5cm]{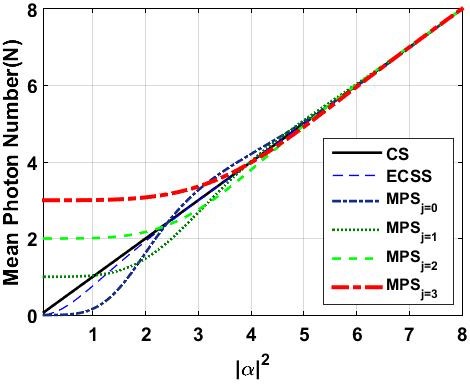}
\caption{\label{fig:2b}The plot shows the variation of mean photon number (N) vs. $|\alpha|^2$. Here, we see that for lower values of $|\alpha|^2$, different states possess different mean photon numbers, while for larger values of $|\alpha|^2 (\geq 5)$ they are the same. Note that, for the very small value of $|\alpha|^2(<<1)$, some states possess 1, 2 and 3 mean photon numbers and in the figure, it seems for $|\alpha|^2=0$.}
\end{figure}

In order to discuss the resolution of MZI for six different states (CS, ECSS, $\text{MPS}_{j=0}$, $\text{MPS}_{j=1}$, $\text{MPS}_{j=2}$ and $\text{MPS}_{j=3}$) we take the variation of $\langle\hat{\Pi}\rangle$ with respect to $\phi$ for different values of $|\alpha|^2$ as shown in Fig. \ref{fig:5(a)}. We can see that the foldness of ECSS, $\text{MPS}_{j=0}$, $\text{MPS}_{j=1}$, $\text{MPS}_{j=2}$ and $\text{MPS}_{j=3}$ increases by one w.r.t. coherent state in the range of $-\pi$ to $\pi$. We can see that the FWHM of the peaks decreases by increasing the photon number. For better understanding, we can see Fig. \ref{fig_6}(a) which shows the variation of FWHM with $|\alpha|^2$. We can see that, $\text{MPS}_{j=0}$ shows the minimum FWHM with respect to other states for the lower values of $|\alpha|^2$. While, for larger values of $|\alpha|^2(>5)$, all the states are giving the same FWHM. Achieving the same FWHM for higher values of $|\alpha|^2$ can be understood by the variation of mean photon number ($N=\langle\hat{a}^{\dagger}\hat{a}\rangle$), which can be written as   
\begin{equation}
\begin{split}
N = N_j\left(X|\alpha|^2-2|\alpha|^2 Y e^{-2|\alpha|^2}\cos(j\pi)\right.\\
+\left. 2|\alpha|^2 V e^{-|\alpha|^2}\cos{\left(|\alpha|^2-\frac{j\pi}{2}+\frac{\pi}{2}\right)}\right).\label{meanP}
\end{split}
\end{equation}
In Fig. \ref{fig:2b}, we plot the $N$ versus $|\alpha|^2$. We can see that for lower values of $|\alpha|^2$, different states possess different mean photon numbers, while for larger values of $|\alpha|^2 (\geq 5)$ they are the same. Note that, for the very small value of $|\alpha|^2(<<1)$, some states possess 1, 2 and 3 mean photon numbers and in the figure, it seems for $|\alpha|^2=0$

To see the effect of photon loss on the FWHM, we plot a graph between FWHM and photon loss (in terms of reflectivity $(r)$ of the fictitious beam splitter shown in Fig. \ref{fig:1}) by considering the mean photon number $N = 3$. In Fig. \ref{fig_6}(b) we can see that, in the lossy case $\text{MPS}_{j=0}$ gives better results than other states. So, we can conclude that $\text{MPS}_{j=0}$ gives better resolution than the other states in both lossless and lossy conditions. 
\begin{figure}
\includegraphics[width=8.5cm, height=11cm]{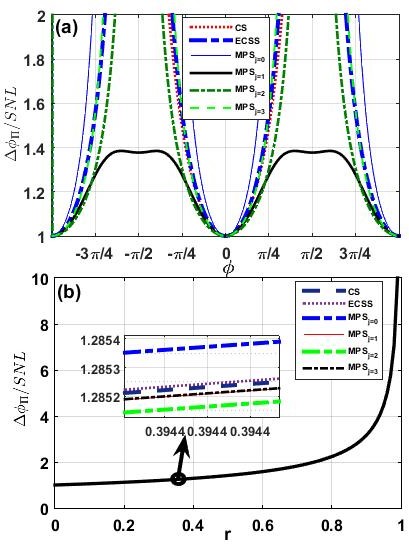}
\caption{\label{fig_8} (a) shows the phase sensitivity of the MZI in parity detection with $|\alpha|^2 = 2$. (b) shows $\Delta\phi/SNL$ vs. r variation with $\phi= 0.02$ and $|\alpha|^2 = 2$.}
 \end{figure}

In order to calculate the phase sensitivity ($\Delta\phi$) of MZI for parity detection, we use the Eq. \eqref{phi}. From Eq. \eqref{32pi}, we can write
\begin{equation}
 \begin{split}
\frac{\partial\langle\hat{\Pi}\rangle}{\partial\phi} =|N_j|^2 e^{-|\alpha|^2}(-2p'X e^{-(2p-|\alpha|^2)}\\+4Vp'\sin(q-p)+2p'Ye^{(2p-|\alpha|^2)}cos(j\pi)).\label{32pid}
 \end{split}   
\end{equation}
Where $p'= \frac{1}{2}|\alpha|^2|t|^2\sin^2\phi,~x'=-\frac{1}{2}|t|^2\sin^2\phi$. So, using Eqs. \eqref{32pi} and \eqref{32pid} in Eq. \eqref{phi} we can calculate the phase sensitivity. 

In order to see the variation in phase sensitivity, we plot the $\Delta\phi/SNL$ versus $\phi$ for all six states in (a) lossless and (b) lossy conditions. From Fig. \ref{fig_8}(a), we can see that all the states saturate the SNL. Note that, except for the better phase point (here, $-\pi$, 0.02 and $\pi$) $\text{MPS}_{j=1}$ (black solid lines) gives the minimum phase sensitivity than the other states in all the regions. That is, in terms of the broader phase range, $\text{MPS}_{j=1}$ performs better as compared to all other states. Further, considering loss (i.e., varying $r$), we plot the $\Delta\phi/SNL$ vs. $r$ by fixing $\phi =0.02$ (Fig. \ref{fig_8}(b)). Under lossy cases, we found that the phase sensitivity of all the states decreases in the same way.

\subsection{$Z$-detection}
For $Z$-detection, we calculate the zero non-zero photon counting probability at the output port. Therefore, from Eq. \eqref{14Z} and \eqref{eq p18}, we can write
\begin{equation}
 \begin{split}
\langle \hat{Z}\rangle=|N_j|^2 (X e^{-p}+2e^{-|\alpha|^2}Vcos(q)\\+Ye^{-|\alpha|^2(1+x)}cos(j\pi)).
\end{split}   \label{eq 20}
\end{equation}
\begin{figure*}
\includegraphics[width= 18.5cm, height=8.5cm]{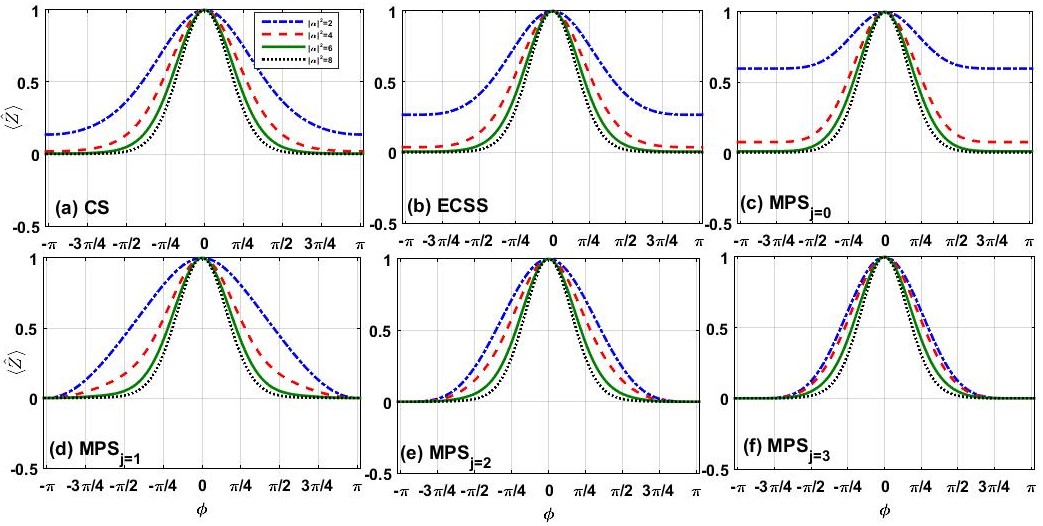}
\caption{\label{fig:5(b)} The figure shows the variation of $\langle\hat{Z}\rangle$ with respect to $\phi$ for different values of $|\alpha|^2$ for the states CS, ECSS, $\text{MPS}_{j=0}$, $\text{MPS}_{j=1}$, $\text{MPS}_{j=2}$ and $\text{MPS}_{j=3}$. We can see that the FWHM of the different states decreases with increasing $|\alpha|^2$.}
\end{figure*}

\begin{figure}[H]
\includegraphics[width=8.5cm, height=10.5cm]{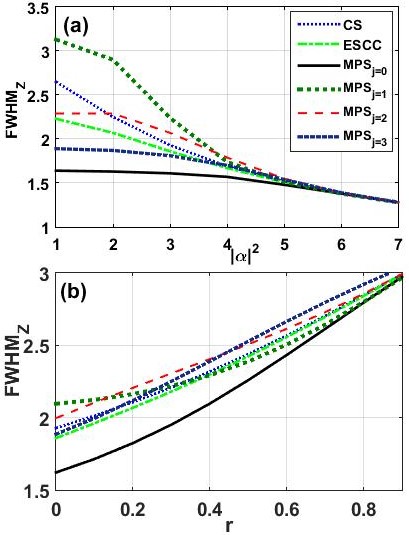}\caption{\label{fig:6}(a) Plot shows the variation of FWHM with $|\alpha|^2$ for the different states in ${Z}$-detection and (b) shows the loss tolerance in ${Z}$-detection with $N = 3$.}
\end{figure}

For the resolution with $Z$-detection, we are plotting $\langle\hat{Z}\rangle$ with respect to $\phi$ by taking different values of $|\alpha|^2$ as shown in Fig. \ref{fig:5(b)}. Here, we observe that the FWHM of the signal decreases by increasing $|\alpha|^2$. Further, for comparing the values of FWHM of all six cases, we plot FWHM with $|\alpha|^2$ as shown in Fig. \ref{fig:6}(a). We can see that, $\text{MPS}_{j=0}$ shows the minimum FWHM with respect to other states. For larger values of $|\alpha|^2(>5)$, all the states give the same FWHM. To see the loss effect on the FWHM, we have plotted a graph between FWHM and $r$ by considering $N = 3$ in Fig. \ref{fig:6}(b). We found that in the lossy case, $\text{MPS}_{j=0}$ performs better than other states. So, we can conclude that $\text{MPS}_{j=0}$ gives better resolution as compared to all other states.

To calculate the phase sensitivity ($\Delta\phi_z$) of MZI for $Z$-detection, we use the Eq. \eqref{15zd}. From Eq. \eqref{eq 20}, we can write
\begin{equation}
 \begin{split}
\frac{\partial\langle \hat{Z}\rangle}{\partial\phi}=|N_j|^2 (-p'X e^{-p}+2p'e^{-|\alpha|^2}V\sin(q)\\-Y|\alpha|^2x'e^{-|\alpha|^2(1+x)}\cos(j\pi)).
\end{split}   \label{eq 20d}
\end{equation}
So, using Eqs. \eqref{eq 20} and \eqref{eq 20d} in Eq. \eqref{15zd} we can calculate the phase sensitivity.

\begin{figure}
\includegraphics[width=8.5cm, height=10.5cm]{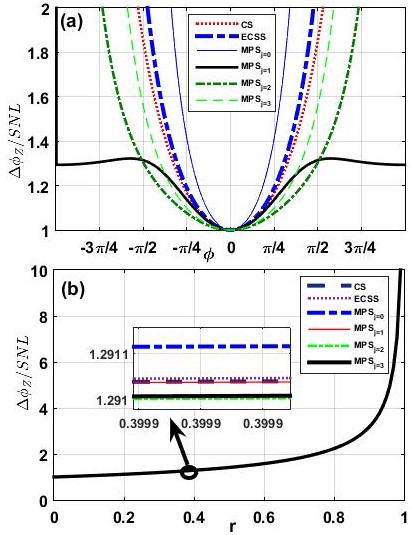}
\caption{\label{fig:8(a)} (a) Shows the phase sensitivity shows $\Delta\phi/SNL$ vs. $\phi$ variation and (b) show $\Delta\phi/SNL$ vs. $r$ with $\phi= 0.02$.}
 \end{figure}

To see the variation in phase sensitivity, we plot the $\Delta\phi_z/SNL$ versus $\phi$ for all six states in both the lossless and lossy conditions. Fig. \ref{fig:8(a)}(a) shows the variations in the lossless case. We can see that all the states saturate SNL but in terms of broader phase range, here, from $-\pi$ to $\pi$, $\text{MPS}_{j=1}$ (black solid lines) gives better results as compared to all other states (Fig. \ref{fig:8(a)}(a)). Further, in the case of photon loss, we plot the $\Delta\phi/SNL$ vs. $r$ at $\phi =0.02$, as shown in Fig. \ref{fig:8(a)}(b). We found that all states behave similarly in lossy conditions.

\section{Observation with MPS and coherent state as inputs}\label{section 5}
In this section, we will discuss the resolution and phase sensitivity of the MZI by using MPS ($|\Psi_j\rangle$) and coherent state ($|\zeta\rangle$) for parity and Z-detection schemes. Our initial state of the system can be written as 
\begin{equation}
|\psi'\rangle_{in}=|\Psi_j,\zeta,0,0\rangle_{a,b,E_a,E_b},\label{eq35}
\end{equation}
where state $|\Psi_j\rangle$ in mode $a$ is given in Eq. \eqref{eq2} and coherent state $|\zeta\rangle$ is in mode $b$. In modes $E_a$ and $E_b$, initially, we have vacuum states because there are no photons in the initial state. Therefore, from Eq. \eqref{3}, the final state of the system at the output of the interferometer is written as

\begin{equation}
     |\psi'\rangle_{out}= N_j\left(A|K'\rangle+B|L'\rangle +C|M'\rangle +D|N'\rangle\right)_{a,b,E_a,E_b}\label{35abcd}
\end{equation},

where, $|K'\rangle,~|L'\rangle,~|M'\rangle$ and $|N'\rangle$ are given in Eq. \eqref{eqB3}. Normalization constant $N_j$ is given in Eq. \eqref{17a} and $A,~B,~C,~D$ are given in Eq. \eqref{eq5}. Eq. \eqref{35abcd} gives the information about the final state of the system which includes the phase change and loss due to the environment. In this section, similar to the previous section, we are investigating the resolution and phase sensitivity of the MZI for parity and $Z$-detection schemes. Expectation values of $\hat{\Pi}$ and $\hat{Z}$ are calculated in terms of the photon number probability as already discussed in Section \ref{section 3}. We have given the detailed expression of the density operator and probability of photon in mode $a$ ($P(n)$) in Appendix \ref{appendix B}.

\subsection{Parity detection}

In parity detection, we can write the expectation value of the parity operator for the system by using Eq. \eqref{eqB7} in Eq. \eqref{11pi}. Therefore the expectation value of the parity operator can be written as
 \begin{widetext}
     \begin{equation}
\begin{split}
\langle\hat{\Pi}\rangle=|N_j|^2(|A|^2(e^{-2G-W})+(|B|^2+|D|^2)e^{-2G}+|C|^2( e^{-2G+W})+e^{-(|\alpha|^2+|\zeta|^2)}(2(|A||B|+|A||D|)(e^{S_1}\cos(T_1-\frac{j\pi}{2}))\\
  +2(|B||C|+|C||D|)(e^{S_2}\cos(T_2 - \frac{j\pi}{2}))+2|B||D|(e^{(S_1-T_1)}\cos(2W-j\pi))+2|A||C|(e^{(S_1-T_1)})\cos(j\pi))).\label{eq38}
\end{split}
\end{equation}
 \end{widetext}

Here,
\begin{equation}
\begin{split}
    G=|\alpha|^2t^2\sin^2(\frac{\phi}{2}) + |\zeta|^2t^2\cos^2(\frac{\phi}{2}),\\
    T_1=O|\alpha|^2-W,~T_2=O|\alpha|^2 + W,\\
    S_1=U|\zeta|^2 - W,~S_2=U|\zeta|^2 + W,\\
    U=(-t^2\cos(\phi) + r^2),~W=t^2|\alpha||\zeta|\sin(\phi),\\
    O=(t^2\cos(\phi) + r^2).\label{eq39}
\end{split}    
\end{equation}
In order to discuss the resolution of the MZI having MPS and coherent states as the inputs, we divide our discussion into two parts. In the first part, we take the lower value of the mean photon ($N=3$) of MPS and vary the mean photon of the coherent state from lower ($|\zeta|^2=3$) to higher values ($|\zeta|^2=100$). In the second case, we vary the mean photon of both MPS and coherent states from lower to higher values.

\begin{figure}
\includegraphics[width=8.5cm, height=5.5cm]{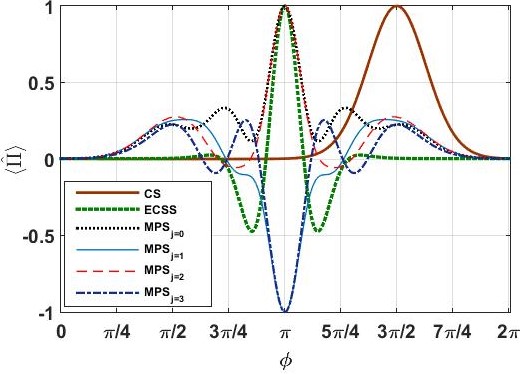}
\caption{\label{fig_19} Here the plots show the variation of all states with high photon numbers ($ N = 3, |\zeta|^2=3$) in the input state in parity detection.}
\end{figure}

\begin{figure*}
\includegraphics[width=18.5cm, height=7.5cm]{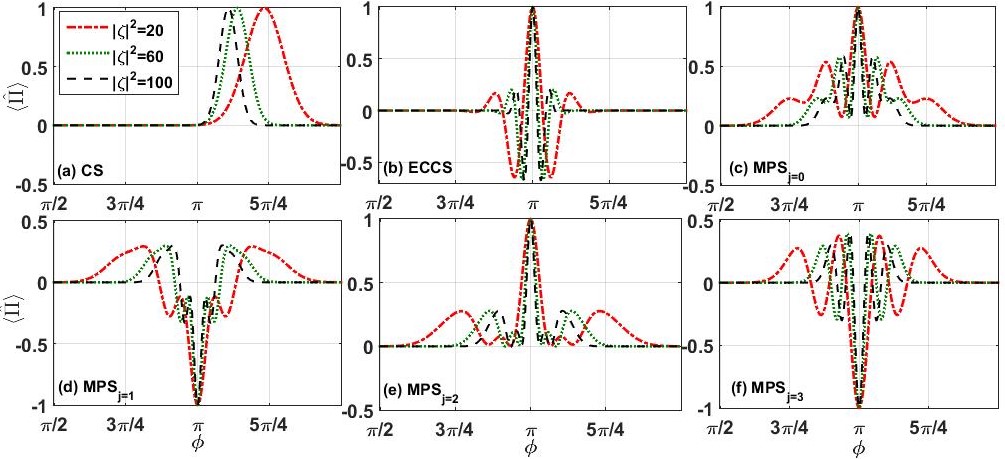}
\caption{\label{fig:12} These plots shows the variation of $\langle\Pi\rangle$ w.r.t. $\phi$ for all six states with varying $|\zeta|^2$ with $N = 3$. On increasing the value of $|\zeta|^2$, we get the decrement in FWHM for all six states}
\end{figure*}

\begin{figure}
\includegraphics[width=8.5cm, height=6cm]{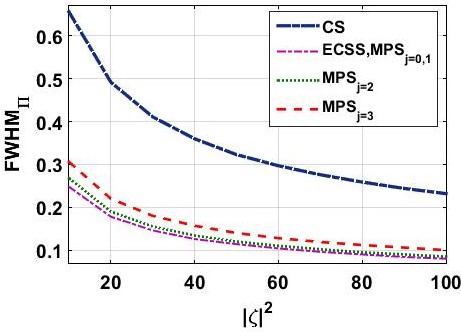}
\caption{\label{fig:11a}The plots show the variation of FWHM of all the six states with $|\zeta|^2$ having $N = 3$.}
\end{figure}

Therefore, initially, we consider $N = 3$ and $|\zeta|^2=3$ and see the variation of $\langle\hat{\Pi}\rangle$ with $\phi$ for all the six states (Fig. \ref{fig_19}). Here, we see the improvement in the foldness of all the five states w.r.t. CS. In which $\text{MPS}_{j=3}$ gives higher foldness than the other four states. If we increase the value of $|\zeta|^2$, we get the decrement in FWHM for all six states, as we can see in Fig. \ref{fig:12}. To see the variation in FWHM with $|\zeta|^2$, we calculated the FWHM for all six states and plotted them in Fig. \ref{fig:11a}. We can see that, $\text{MPS}_{j=0}$, $\text{MPS}_{j=1}$ and ECSS gives minimum FWHM as compared to the other three states with increasing $|\zeta|^2$ (Fig. \ref{fig:11a}). From Fig. \ref{fig_6}(a) and Fig. \ref{fig:11a}, we can say that MZI with MPS as one of the inputs having lower energy and another input port injected with a coherent state enhances the resolution of the MZI.

\begin{figure*}
\includegraphics[width=18cm, height=9cm]{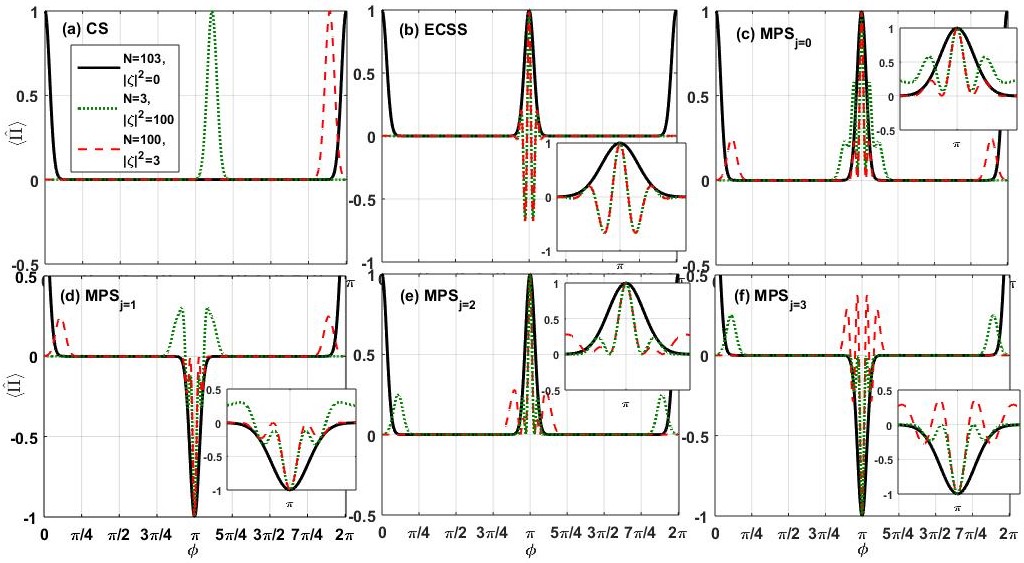}
\caption{\label{fig:16} These plots show the variation of all states with high photon numbers in the input state in parity detection.}
\end{figure*}

\begin{figure}
\includegraphics[width=8.5cm, height=6.5cm]{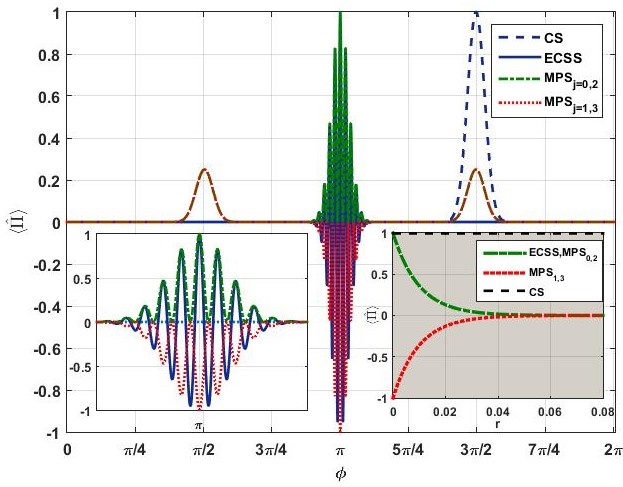}
\caption{\label{fig:17}Plots show the variation of $\langle\hat{\Pi}\rangle$ versus $\phi$ for all six states with $N=51$ $|\zeta|^2=52$. As we can see if we put a nearly equal number of photos in coherent state and MPS then the foldness of the output signal increases as compared to MPS with a vacuum as an input. We can see that $\text{MPS}_{j=0,1,2,3}$ gives extra peaks in compare to ECSS.}
\end{figure}

Now, we are interested in varying both $N$ and $|\zeta|^2$, i.e., the energy variation in both inputs of the MZI. In order to discuss this, in detail, we divide it into four cases. (i) $N=103$ and $|\zeta|^2=0$; (ii) $N=3$ and $|\zeta|^2=100$; (iii) $N=100$ and $|\zeta|^2=3$; and (iv) $N=51$ and $|\zeta|^2=52$. Here, we ensure that the energy inputs are consistent across all four cases. In Fig. \ref{fig:16}, we can see that on varying the values of $N$ and $|\zeta|^2$, foldness in the resolution increases in all five states as compared to the CS. For the fourth case (i.e., $N=51$ and $|\zeta|^2=52$), we are getting the highest foldness as compared to the other three cases (Fig. \ref{fig:17}). Here, it is important to note that, $\text{MPS}_{j=0}$, $\text{MPS}_{j=1}$, $\text{MPS}_{j=2}$ and $\text{MPS}_{j=3}$ give two extra peaks  at $\pi/2$ and $3\pi/2$ as compared to ECSS case (Fig. \ref{fig:17}). As we can see ECSS and MPSs give the peak count 10 $\approx \sqrt{103}$ (we count only lower or upper peaks) in the mid-region. This is in agreement with the Wang \textit{et al.} results \cite{wang2016super} that foldness increases with the square root of the input mean photon number. So in this way, ECSS and MPSs provide us with the same number of peaks in that region.

This means we are getting extra folding in the resolution for the case of $\text{MPS}_{j=0,1,2,3}$ in comparison to the case of ECSS. Thus, we can conclude that an equal photon number in both input ports gives a better resolution than an unequal one.

In order to see the effect of loss on the foldness we consider $N=51$ and $|\zeta|^2=52$ case. We plot the variation of $\langle\hat{\Pi}\rangle$ with $r$ (grey inset of the Fig. \ref{fig:17}) with the optimal value of $\phi=\pi$. We can see that for $> 4 \%$ loss, we lose the foldness for all five states. While, further, we found that the foldness at $\pi/2$ and $3\pi/2$ for $\text{MPS}_{j=0,1,2,3}$ remains static and FWHM increases with loss. One more point here we mention that, in our analysis, we found that the loss tolerance of foldness decreases with increasing photon number in the state.

\begin{figure}
\includegraphics[width=8.5cm, height=10cm]{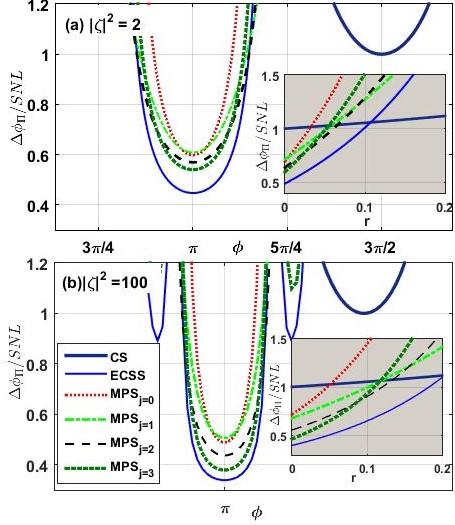}
\caption{\label{fig_14} These two plots show the variation of $\Delta\phi$ versus $\phi$ plot for all six chosen state. (a) shows the variation with $|\zeta|^2=2$ and $|\alpha|^2=2$, (b) shows the variation with $|\zeta|^2 = 100$ and $|\alpha|^2=2$.}
\end{figure}

\begin{figure}
\includegraphics[width=8.5cm, height=5.5cm]{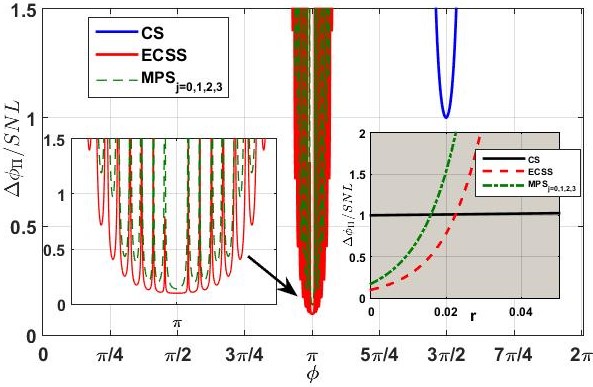}
\caption{\label{fig_18new}Graphs show the variation of $\Delta\phi/SNL$ versus $\phi$ for all six states with $N =51$ $|\zeta|^2=52$. We can see that for ECSS, $\text{MPS}_{j=0,1,2,3}$, phase sensitivity beats the SNL. In the grey inset, we show the effect of loss on the phase sensitivity.}
\end{figure}

In order to calculate the phase sensitivity ($\Delta\phi$) of MZI for parity detection, we use the Eq. \eqref{phi}. From Eq. \eqref{eq38}, we can write
\begin{widetext}
     \begin{equation}
\begin{split}
\frac{\partial\langle\hat{\Pi}\rangle}{\partial\phi}=|N_j|^2\left(|A|^2e^{-2G-W}(-2G'-W')+(|B|^2+|D|^2)e^{-2G}(-2G')+|C|^2e^{-2G+W}(-2G'+W')\right.\\
+e^{-(|\alpha|^2+|\zeta|^2)}(2(|A||B|+|A||D|)(e^{S_1}S_1'\cos(T_1-\frac{j\pi}{2})-e^{S_1}T_1'\sin(T_1-\frac{j\pi}{2}))\\
  +2(|B||C|+|C||D|)(e^{S_2}S_2'\cos(T_2 - \frac{j\pi}{2})-e^{S_2}T_2'\sin(T_2 - \frac{j\pi}{2}))\\
  \left.+2|B||D|(e^{(S_1-T_1)}(S_1'-T_1')\cos(2W-j\pi)-2W'e^{(S_1-T_1)}\sin(2W-j\pi))+2|A||C|e^{(S_1-T_1)}(S_1'-T_1')\cos(j\pi))\right).
\end{split}
\end{equation}
 \end{widetext}
Here, $G,~W,~S_1,~S_2,~T_1$ are given in Eq. \eqref{eq39} and
\begin{equation}
\begin{split}
    G'=\frac{1}{2}(|\alpha|^2-|\zeta|^2)t^2\sin(\phi),~T_1'=O'|\alpha|^2-W',\\
    T_2'=O'|\alpha|^2 + W',~S_1'=U'|\zeta|^2 - W',\\
    S_2'=U'|\zeta|^2 + W',~U'=t^2\sin(\phi),\\
    W'=t^2|\alpha||\zeta|\cos(\phi),~O'=-t^2\cos(\phi).\label{eq391}
\end{split}    
\end{equation}
So, using Eqs. \eqref{eq38} and \eqref{eq391} in Eq. \eqref{phi} we can calculate the phase sensitivity.
For the phase sensitivity, we see the variation $\Delta\phi/SNL$ with $\phi$. Similar to the resolution part, we will discuss the results for different input photon variations. Firstly, we consider two cases: (i) for  $|\zeta|^2 = 2$, $|\alpha|^2 = 2$ and, (ii) for  $|\zeta|^2 = 100$, $|\alpha|^2 = 2$. In both cases, we have found that $\Delta\phi/SNL$ beats the SNL for all the states except the CS (Fig. \ref{fig_14}). This shows that MPS with CS ($|\zeta\rangle$) as the input of the MZI gives super phase sensitivity. Also, note that with a high photon number ($|\zeta|^2=100$) in the coherent state, we get the improvement in the phase sensitivity (Fig. \ref{fig_14}). In the photon loss, for $|\zeta|^2 = 2$, $|\alpha|^2 = 2$, we found that ECSS gives approximately $10\%$ and the other four give approximately $5\%$ photon loss (grey inset in Fig. \ref{fig_14}(a)), while in $|\zeta|^2 = 100$, $|\alpha|^2 = 2$, ECSS gives approximately $20\%$ and the other four give approximately $10\%$ photon loss (grey inset in Fig. \ref{fig_14}(b)). Now, we consider the case in which $|\alpha|^2=51$, $|\zeta|^2=52$. We get a significant change in the phase sensitivity, as shown in Fig. \ref{fig_18new}. But in this case, we found a very small loss tolerance (grey inset in Fig. \ref{fig_18new}).

\subsection{Z-detection}
For $Z$-detection, we calculate the zero non-zero photon counting probability at the output port. Therefore, from Eq. \eqref{14Z} and \eqref{B6}, we can write
\begin{widetext}
\begin{equation}
 \begin{split}
\langle \hat{Z}\rangle=|N_j|^2((|A|^2e^{(-G-W)} + (|B|^2+ |D|^2)e^{-G}
+|C|^2e^{(-G+W)})+2e^{-(|\alpha|^2+|\zeta|^2)}((|A||B|\\
+|A||D|)e^{(\frac{1}{2}(S_1)+\frac{1}{2}|\zeta|^2)}\cos(\frac{1}{2}(T_1+|\alpha|^2)-\frac{j\pi}{2})
+(|B||C|+|C||D|)e^{(\frac{1}{2}(S_2)+\frac{1}{2}|\zeta|^2)}\cos(\frac{1}{2}(T_2+|\alpha|^2)-\frac{j\pi}{2})\\
+|B||D|e^{(\frac{1}{2}(S_1-T_1)+\frac{1}{2}(|\zeta|^2-|\alpha|^2))}\cos(j\pi-W)
+|A||C|e^{(\frac{1}{2}(S_1-T_1)+\frac{1}{2}(|\zeta|^2-|\alpha|^2))}\cos({j\pi})))\label{pcm}
\end{split}   
\end{equation}
\end{widetext}

\begin{figure}
\includegraphics[width=8.5cm, height=5.5cm]{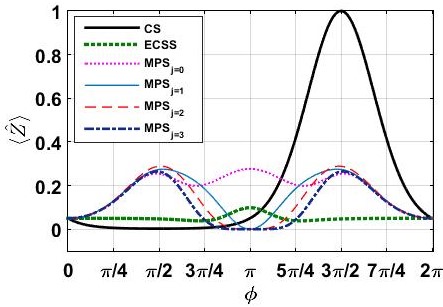}
\caption{\label{fig_20a}Plots show the variation of $\langle\hat{Z}\rangle$ versus $\phi$ with $|\zeta|^2 =3$ and $N=3$. We can see that $\text{MPS}_{j=0,1,2,3}$ give foldness greater than the CS, and ECSS}
\end{figure}

\begin{figure*}
\includegraphics[width=18.5cm, height=7.5cm]{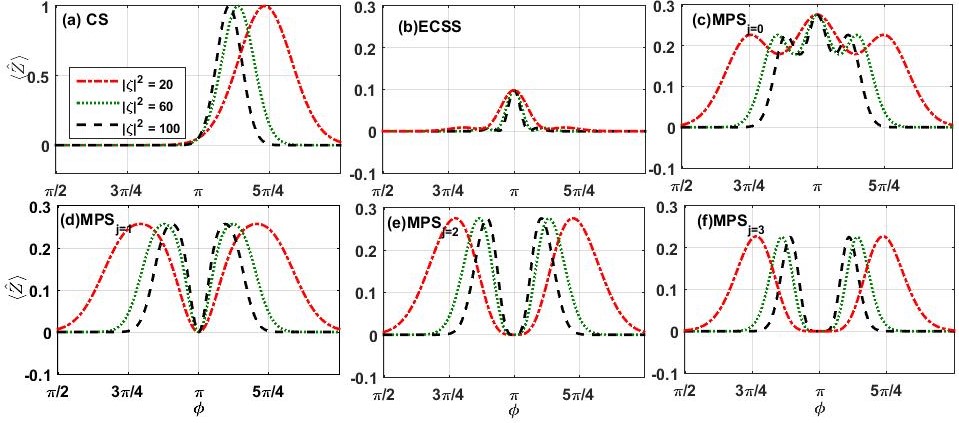}
\caption{\label{fig:13} These plots show the variation of $\langle Z\rangle$ w.r.t. $\phi$ for all six states with varying $|\zeta|^2$ and $N = 3$. We can see that on increasing the value of $|\zeta|^2$, FWHM for all the six states decrease.}
\end{figure*}

So, with the help of $\langle\hat{Z}\rangle$, we calculate the resolution and phase sensitivity. In order to discuss the resolution for all the six states with CS ($|\zeta\rangle$) as the second input of the interferometer, we plotted $\langle\hat{Z}\rangle$ versus $\phi$. Firstly, we consider the $N = 3$, $|\zeta|^2=3$ and we find that $\text{MPS}_{j=0,1,2,3}$ give foldness greater than the CS, and ECSS, as shown in Fig. \ref{fig_20a}. As we increase the value of $|\zeta|^2$, FWHM for all the six states decreases (Fig. \ref{fig:13}).

\begin{figure*}
\includegraphics[width=18.5cm, height=8.5cm]{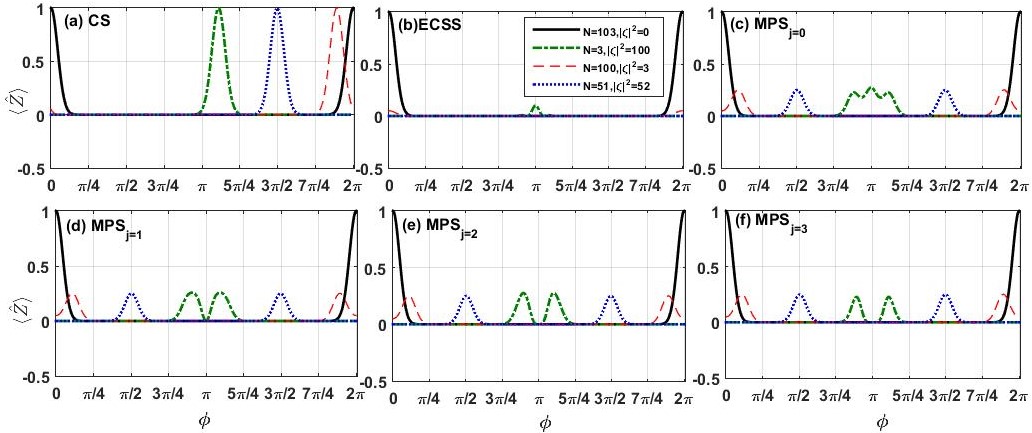}
\caption{\label{fig:18ab} These plots show the variation of all six states with high photon numbers in the input state. Here all four MPS show increased foldness but the peak height is too small. We can see that, in this case, the performance of ECSS is very poor.}
\end{figure*}

\begin{figure*}
\includegraphics[width=18.5cm, height=8cm]{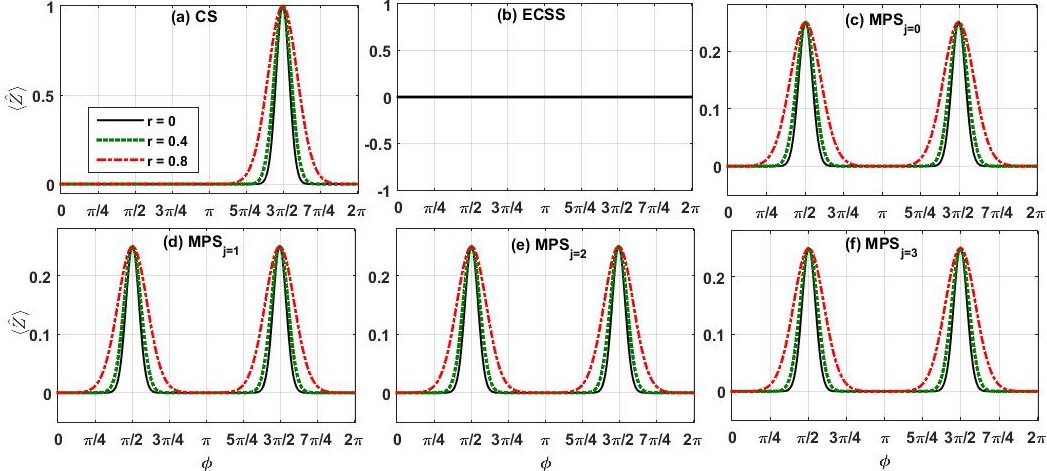}
\caption{\label{fig:21} These plots show the variation of all six states with $N = 51$, $|\zeta|^2 = 52$ in the input state in lossy condition. Here all four MPS show 2-fold peaks than CS in lossless ($ r=0 $) as well as in lossy ($ r\neq 0 $) conditions. The performance of ECSS is very poor also in this case.}
\end{figure*}

Now, we are interested in varying the values of $|\alpha|^2$ and $|\zeta|^2$ to see the effect on the resolution of the MZI. In Fig. \ref{fig:18ab}, we can see the plots of $\langle\hat{Z}\rangle$ versus $\phi$ for four different cases of mean photon number ($N$) and $|\zeta|^2$. From Fig. \ref{fig:18ab}, we can easily see that $\text{MPS}_{j=0}$, $\text{MPS}_{j=1}$, $\text{MPS}_{j=2}$ and $\text{MPS}_{j=3}$ performs better than ECSS and CS in all the situation except when $N=103$ and $|\zeta|^2=0$. Here, we notice that as we increase the value of $N$ we found that, for higher values, $\text{MPS}_{j=0,1,2,3}$ gives approximately the same results. Here for the case of $N = 51$ and $|\zeta|^2 = 52$, we found that $\text{MPS}_{j=0,1,2,3}$ gives 2 fold resolution as compared to CS in loss also (Fig. \ref{fig:21}). ECSS performs worse in lossless as well as in lossy conditions (Fig. \ref{fig:21}).  So, in conclusion, we can say that $\text{MPS}_{j=0,1,2,3}$ gives better results than ECSS and CS for $Z$-detection. Since $Z$-detection is more feasible than parity detection, we can say that with $\text{MPS}$ we are getting better results in comparison to parity detection with CS as one of the inputs of MZI. 

\begin{figure}
\includegraphics[width=8.5cm, height=10.5cm]{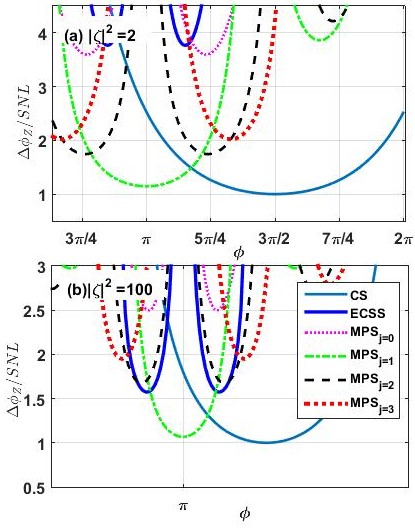}
\caption{\label{fig:14} Plots show the variation of $\Delta\phi/SNL$ versus $\phi$ for all six states. (a) Shows the variation with $|\alpha|^2=2$, $|\zeta|^2=2$, and (b) shows the variation with $|\alpha|^2=2$, $|\zeta|^2 = 100$.}
\end{figure}

In order to calculate the phase sensitivity ($\Delta\phi$) of MZI for Z-detection, we use the Eq. \eqref{15zd}. From Eq. \eqref{pcm}, we can write
\begin{widetext}
\begin{equation}
 \begin{split}
\frac{\partial \langle \hat{Z}\rangle}{\partial\phi}=|N_j|^2((|A|^2e^{(-G-W)}(-G'-W') + (|B|^2+ |D|^2)e^{-G}(-G')
+|C|^2e^{(-G+W)}(-G'+W'))\\
+2e^{-(|\alpha|^2+|\zeta|^2)}((|A||B|+|A||D|)(\frac{1}{2}S_1'e^{(\frac{1}{2}S_1+\frac{1}{2}|\zeta|^2)}\cos(\frac{1}{2}(T_1+|\alpha|^2)-\frac{j\pi}{2})-\frac{1}{2}T_1'e^{(\frac{1}{2}S_1+\frac{1}{2}|\zeta|^2)}\sin(\frac{1}{2}(T_1+|\alpha|^2)-\frac{j\pi}{2}))
\\
+(|B||C|+|C||D|)(\frac{1}{2}S_2'e^{(\frac{1}{2}S_2+\frac{1}{2}|\zeta|^2)}\cos(\frac{1}{2}(T_2+|\alpha|^2)-\frac{j\pi}{2})-\frac{1}{2}T_2'e^{(\frac{1}{2}(S_2)+\frac{1}{2}|\zeta|^2)}\sin(\frac{1}{2}(T_2+|\alpha|^2)-\frac{j\pi}{2}))\\
+|B||D|(\frac{1}{2}(S_1'-T_1')e^{(\frac{1}{2}(S_1-T_1)+\frac{1}{2}(|\zeta|^2-|\alpha|^2))}\cos(j\pi-W)+W'e^{(\frac{1}{2}(S_1-T_1)+\frac{1}{2}(|\zeta|^2-|\alpha|^2))}\sin(j\pi-W))\\
+\frac{1}{2}(S_1'-T_1')|A||C|e^{(\frac{1}{2}(S_1-T_1)+\frac{1}{2}(|\zeta|^2-|\alpha|^2))}\cos({j\pi}))). \label{PLJ}
\end{split}
\end{equation}
\end{widetext}
In order to discuss the result, we consider two cases: (i) $|\alpha|^2=2$, $|\zeta|^2=2$; (ii) $|\alpha|^2=2$, $|\zeta|^2=100$. We plot the $\Delta\phi/SNL$ versus $\phi$ for both the cases (Fig. \ref{fig:14}). From Fig. \ref{fig:14}, we can see that we get better phase sensitivity in both cases for CS and $\text{MPS}_{j=1}$.

\section{discussion and conclusions}\label{section 6}

In this paper, we have investigated the enhancement in phase sensitivity and resolution in MZI-based quantum LiDAR by employing MPS with vacuum state and MPS with coherent state ($|\zeta\rangle$) as the inputs by party and $Z$-detection as measurement schemes.

In Section \ref{section 4}, we have discussed the MPS with Vacuum state case. In this case, we found that $\text{MPS}_{j=0}$ gives better resolution in both detection scenarios compared to all other states in lossless (Fig. \ref{fig_6}(a) and Fig. \ref{fig:6}(a)) as well as in lossy conditions (Fig. \ref{fig_6}(b) and Fig. \ref{fig:6}(b)). All states reached the standard quantum limit (SNL) regarding phase sensitivity in both detection cases. However, for a wide range of phase values, $\text{MPS}_{j=1}$ exhibited enhanced performance (Fig. \ref{fig_8}(a) and Fig. \ref{fig:8(a)}(a)). In lossy conditions, all states degraded in the same way (Fig. \ref{fig_8}(b) and Fig. \ref{fig:8(a)}(b)).

In Section \ref{section 5}, we have discussed the MPS with a coherent state case. We found that when both inputs contain an equal mean number of photons, foldness in resolution increases with $\approx \sqrt{N_T}$, where $N_T$ is the total mean number of photons in the inputs. For ECSS with the coherent state as the inputs, this result was also shown by the Wang \textit{et al.} in \cite{wang2016super}. However, here, in the MPS case (i.e. for $\text{MPS}_{j=0,1,2,3}$), we get a pair of peaks (at $\pi/2$ and $3\pi/2$) along with $\sqrt{N_T}$ foldness improvement (Fig. \ref{fig:17}). The important thing is that in the lossy case, we lose the improved $\sqrt{N_T}$ foldness (grey inset in Fig. \ref{fig:17}). However, the loss does not affect the extra pair of peaks for MPS. In conclusion in lossy condition, we still have improved fondness compared to the CS and ECSS cases. Note that, the emergence of an extra pair not only happens in equal input cases but also happens when we take unequal input cases, for example in Fig. \ref{fig:16}, for $N=3, |\zeta|^2=100$, we get the extra peaks for $\text{MPS}_{j=2,3}$ and for $N=100, |\zeta|^2=3$, we get the extra peaks for $\text{MPS}_{j=0,1}$. Concerning phase sensitivity, all states except the coherent state surpassed the SNL (Fig. \ref{fig_14}). In terms of loss tolerance, for low photon numbers in CS, the ECSS exhibited a loss tolerance of less than 10\%. At the same time, $\text{MPS}_{j=0,1,2,3}$ states showed a loss tolerance of less than 5\% compared to the coherent state (grey inset in Fig. \ref{fig_14}(a)), While for higher photon numbers in CS, ECSS shows $20\%$ and $\text{MPS}_{j=0,1,2,3}$ shows approximately $10\%$ photon loss (grey inset in Fig. \ref{fig_14}(b)).  However, with higher photon numbers, loss tolerance decreased (grey inset of Fig. \ref{fig_18new}). 

In the case of the Z-detection scheme, $\text{MPS}_{j=0,1,2,3}$ outperformed the ECSS and coherent state in terms of resolution (Fig. \ref{fig_20a}, Fig. \ref{fig:13}, and Fig. \ref{fig:18ab}).  For higher and equal numbers of photons in the input state we found that $\text{MPS}_{j=0,1,2,3}$ gives 2 fold resolution as compared to CS in lossless as well as lossy conditions (Fig. \ref{fig:21}). $\text{MPS}_{j=0,1,2,3}$ gives better results than ECSS and CS for $Z$-detection. As for phase sensitivity, the coherent state and $\text{MPS}_{j=1}$ yielded better results (Fig. \ref{fig:14}).

 In summary, the four multi-photon states ($\text{MPS}_{j=0,1,2,3}$), introduced by Mishra et al. \cite{mishra2021ququats}, exhibit super-resolution and super-phase sensitivity in some conditions. Since with MPS, we are getting super sensitivity, so, we can get improvement in the ranging measurements also. As we know the range of the target $(2f)$ is related to the phase  $\phi$ with the relation $\phi= 2kf$, where $k = \frac{2\pi}{\lambda}$ is the wave number and $\lambda$ is the wavelength of the light. From here it is clear that $\Delta\phi \propto \Delta f$ means better measurement in phase sensitivity improves the raging performance. We also get the improvement in super-resolution and it will improve the image quality. Because resolution is the key factor in image formation \cite{blakey2022quantum, liu2023compact, dowling2014quantum, malik2014quantum, PhysRevA.78.063828}. The root cause of the outstanding performance of MPS is the non-classicality which we have shown by the comparisons of Wigner functions of all states in Fig. \ref{fig:2(C)}. Property of the non-classicality comes from the bipartite four-component entangled coherent states \cite{mishra2021ququats}. Which results from the interaction of the MPSs with a 50:50 beam splitter.

These states outperform the coherent state in most scenarios and also surpass the even coherent superposition state (ECSS) \cite{wang2016super, qiang2018effects, WANG20163717} in many cases. The experimental scheme for generating these multi-photon states has been proposed in \cite{mishra2021ququats}. Consequently, multi-photon states hold promise as an alternative nonclassical resource for quantum imaging and quantum sensing applications, such as in quantum LiDAR.

\begin{acknowledgments}
PS acknowledges UGC for the UGC Research Fellowship. PS would like to thank Mr. Gaurav Shukla for his useful comments and suggestions. DKM acknowledges financial support from the Science \& Engineering Research Board (SERB), New Delhi for CRG Grant (CRG/2021/005917) and Incentive Grant under Institution of Eminence (IoE), Banaras Hindu University, Varanasi, India. MKM would like to thank Shri Nilesh M. Desai (Director, SAC, Ahmedabad) and Dr. Rashmi Sharma for their encouragement and support.  
\end{acknowledgments}

\begin{widetext}
\appendix
\section{Calculation of Wigner function}\label{appendix wigner}
The Wigner function of a state with density operator $\hat{\rho}$ is written in Eq. (\ref{wf7}). Density operator of state $|\Psi_j\rangle$ is written as $|\Psi_j\rangle\langle\Psi_j|$. Therefore,  
\begin{equation}
   \text{Tr}[\hat{\rho}\hat{D}(\eta)] =  \langle \Psi_j|\exp{(\eta\hat{a}^\dagger-\eta^*\hat{a})}|\Psi_j\rangle.
\end{equation}
On solving the above trace and putting this in Eq. (\ref{wf7}), we get the Wigner function for the MPS
    \begin{equation}
 \begin{split}
      W_j(\lambda)= \frac{2N_j}{\pi}\left(e^{-2(|\alpha|^2+|\lambda|^2)}(|A|^2e^{4p_1}+|B|^2e^{4q_1}+|C|^2e^{-4p_1}+|D|^2e^{-4q_1})+2e^{(2(p_1-q_1)-2|\lambda|^2-|\alpha|^2)}(|A||B|\cos(2(p_1-q_1)
      \right.\\
      +|\alpha|^2+\frac{j\pi}{2}) +|A||D|\cos(-2(p_1-q_1) +|\alpha|^2+\frac{j\pi}{2}))+2e^{-2|\lambda|^2}(|A||C|\cos(4q_1-j\pi )+|B||D|\cos(4p_1-j\pi))\\
      +2|C||D|e^{(2(p_1-q_1)-4p_1-2|\lambda|^2-|\alpha|^2)}\cos(2(p_1-q_1)-4p_1-|\alpha|^2+\frac{j\pi}{2})\\
     \left.+2|B||C|e^{(-2(p_1-q_1)-2|\lambda|^2-|\alpha|^2)}\cos(2(p_1-q_1)+|\alpha|^2+\frac{j\pi}{2})\right),
 \end{split}\label{a45}
\end{equation}
where,
$\alpha=x_1+ix_2,~\lambda=y_1+iy_2,~p_1 =x_1y_1+x_2y_2,~q_1 =x_1y_2-x_2y_1.$
This is a generalised expression of the Wigner function, so, by putting appropriate values of $A, B, C, D$ and $j$ we can write the Wigner function for four MPSs and for CS and ECSS. For example, the Wigner function for coherent state $|\xi\rangle$ can be written (by putting $A=1,~B=C=D=j=0$) as 
\begin{equation}
\begin{split}
 W(\lambda)= \frac{1}{\pi^2}\int d^2\eta e^{(\frac{-|\eta|^2}{2}+\eta \xi^*-\eta^*\xi)}e^{-(\eta\lambda^{*}-\eta^{*}\lambda)},
 \end{split}
 \end{equation}
is the wigner function for the coherent state.
 
\section{Calculation with MPS and vacuum state as inputs}\label{appendix a}
\subsection{Calculation of density operator}
The final state of the system at the output of MZI is
\begin{equation}
\begin{split}
     |\psi\rangle_{out}= N_j\left[A|K\rangle
     +B|L\rangle +C|M\rangle
+D|N\rangle\right]_{a,b,E_a,E_b},
\end{split}
\end{equation}
where, $N_j$ is given in Eq. \eqref{17a}. $A,~B,~C,~D$ are given in Eq. \eqref{eq5} and $|K\rangle,~|L\rangle,~|M\rangle$ and $|N\rangle$ are given as
\begin{equation}
        \begin{split}
        |K\rangle=|\alpha\vartheta, \alpha\varsigma,{\Bar{r}\alpha e^{i\phi}}, {i\Bar{r}\alpha}\rangle,
        |L\rangle=|\beta\vartheta, \beta\varsigma, {\Bar{r}\beta e^{i\phi}}, {i\Bar{r}\beta}\rangle,\\
        |M\rangle=|\gamma \vartheta, \gamma\varsigma, {\Bar{r}\gamma e^{i\phi}}, {i\Bar{r}\gamma}\rangle,
        |N\rangle=|\delta\vartheta, \delta\varsigma, {\Bar{r}\delta e^{i\phi}}, {i\Bar{r}\delta}\rangle.\label{eqA3}
        \end{split}
\end{equation}
where, $\vartheta=ite^{\frac{i\phi}{2}}\sin\left(\frac{\phi}{2}\right)$, $\varsigma=ite^{\frac{i\phi}{2}}\cos\left(\frac{\phi}{2}\right)$ and $\Bar{r}=ir/\sqrt{2}$.

The density operator ($\hat{\rho}_{out}$) for the final state of the system is written as
\begin{equation}
\begin{split}
\hat{\rho}_{out}= |\psi\rangle_{out~out}\langle\psi|
=|N_j|^2\left(|A|^2|K\rangle\langle K|\right.
+|B|^2|L\rangle\langle L|+|C|^2|M\rangle\langle M|+|D|^2|N\rangle\langle N|\\
+AB^*|K\rangle\langle L|+BA^*|L\rangle\langle K|+AC^*|K\rangle\langle M|
+CA^*|M\rangle\langle K|+AD^*|K\rangle\langle N|+DA^*|N\rangle\langle K|\\
+BC^*|L\rangle\langle M|+CB^*|M\rangle\langle L|+BD^*|L\rangle\langle N|
+\left.DB^*|N\rangle\langle L|+CD^*|M\rangle\langle N|+DC^*|N\rangle\langle M|\right).
\end{split}\label{221abc}
\end{equation}

\subsection{Calculation of $P(\pm)$}\label{appendix a2}
In parity detection, the parity operator, $\hat{\Pi}$, divides the photon counting data $\{n,m\}$ into binary outcomes $\pm$, according to the even or odd number of photons at the output port. For even number of photons, from Eq. \ref{eq p18}, we write the probability as
\begin{equation}
 \begin{split}
P(+) =|N_j|^2e^{-|\alpha|^2} \left(Xe^{-p + |\alpha|^2}\left(\frac{1}{2}(e^p + e^{-p})\right)
+V\left(e^{-iq}\frac{1}{2}(e^{-ip} + e^{ip})+c.c.\right)+Y\left(e^{q'}\frac{1}{2}(e^{-p} + e^{p})+c.c.\right)\right)
 \end{split} 
\end{equation}
Similarly for odd number of photons
\begin{equation}
 \begin{split}
P(-) =|N_j|^2e^{-|\alpha|^2} \left(Xe^{-p + |\alpha|^2}\left(\frac{1}{2}(e^p - e^{-p})\right)
+V\left(e^{-iq}\frac{1}{2}(e^{-ip} - e^{ip})+c.c.\right)+Y\left(e^{q'}\frac{1}{2}(e^{-p} - e^{p})+c.c.\right)\right)
 \end{split} 
\end{equation}

\section{Calculation with MPS and coherent state as inputs}\label{appendix B}
The final state of the system at the output of the interferometer can be written as
\begin{equation}
     |\psi'\rangle_{out}= N_j\left(A|K'\rangle+B|L'\rangle +C|M'\rangle +D|N'\rangle\right)_{a,b,E_a,E_b},
\end{equation}
with, $N_j$ is given in Eq. \eqref{17a} and $A,~B,~C,~D$ are given in Eq. \eqref{eq5} and $|K'\rangle,~|L'\rangle,~|M'\rangle$ and $|N'\rangle$ are given as
\begin{equation}
    \begin{split}
        |K'\rangle=|k_1',k_2',k_3',k_4'\rangle,
        |L'\rangle=|l_1',l_2',l_3',l_4'\rangle,\\
        |M'\rangle=|m_1',m_2',m_3',m_4'\rangle,
        |N'\rangle=|n_1',n_2',n_3',n_4'\rangle.
    \end{split}\label{eqB3}
\end{equation}
Where,

\begin{equation}
    \begin{split}
k_1'= k_1+\varsigma',~k_2'= k_2+\vartheta',~k_3'=k_3+\Bar{r}',k_4'={k_4+\Bar{r}\zeta},\\~l_1'= l_1+\varsigma',~l_2'= l_2+\vartheta',~l_3^{'}=l_3+\Bar{r}',~l_4'=l_4+\Bar{r}\zeta,\\~m_1'= m_1+\varsigma',~m_2'= m_2+\vartheta',~m_3^{'}=m_3+\Bar{r}',\\
~m_4'=m_4+\Bar{r}\zeta,~n_1'= n_1+\varsigma',
~n_2'=n_2+\vartheta',\\
n_3'=n_3+\Bar{r}',~n_4'=n_4+\Bar{r}\zeta,~\vartheta'=-it\zeta e^{\frac{i\phi}{2}}\sin\left(\frac{\phi}{2}\right),\\
\varsigma'=it\zeta e^{\frac{i\phi}{2}}\cos\left(\frac{\phi}{2}\right),~\Bar{r}=ir/\sqrt{2},~\Bar{r}'=-re^{i\phi}\zeta/\sqrt{2}.
    \end{split}\label{eqB4}
\end{equation}

The density operator $\hat{\rho}'_{out}$ is written as
\begin{equation}
\begin{split}
\hat{\rho}'_{out}= |\psi'\rangle_{out~out}\langle\psi'|=|N_j|^2(|A|^2|K'\rangle\langle K'|
+|B|^2|L'\rangle\langle L'|+|C|^2|M'\rangle\langle M'| +|D|^2|N'\rangle\langle N'|\\
+AB^{*}|K'\rangle\langle L'|+BA^{*}|L'\rangle\langle K'| +AC^{*}|K'\rangle\langle M'|
+CA^{*}|M'\rangle\langle K'|+AD^{*}|K'\rangle\langle N'| +DA^{*}|N'\rangle\langle K'|\\
+BC^{*}|L'\rangle\langle M'|+CB^{*}|M'\rangle\langle L'| +BD^{*}|L'\rangle\langle N'|
+DB^{*}|N'\rangle\langle L'|+CD^{*}|M'\rangle\langle N'| +DC^{*}|N'\rangle\langle M'|).\label{22ab1}
\end{split}
\end{equation}
Where $A,~B,~C,~D$ are given in Section \eqref{eq5}. 

Now, from Eq. \eqref{p9}, the probability of getting $n$ photon at port $a$ can be written as
     \begin{equation}
\begin{split}
  P(n)=|N_j|^2\left(|A|^2 e^{-|k_1'|^2}\frac{(|k_1'|^2)^n}{n!}+|B|^2 e^{-|l_1'|^2}\frac{(|l_1'|^2)^n}{n!}+|C|^2e^{-|m_1'|^2}\frac{(|m_1'|^2)^n}{n!} + |D|^2 e^{-|n_1'|^2}\frac{(|n_1'|^2)^n}{n!}\right.\\
  +e^{-|\alpha|^2}\left(AB^* e^{k'_2l_2^{'*}+k'_3l_3^{'*}+k'_4l_4^{'*}}\frac{(k'_1l_1^{'*})^n}{n!} + A^*B e^{k_2^{'*}l'_2+k_3^{'*}l'_3+k_4^{'*}l'_4}\frac{(k_1^{'*}l'_1)^n}{n!} + AC^* e^{k'_2m_2^{'*}+k'_3m_3^{'*}+k'_4m_4^{'*}}\frac{(k'_1m_1^{'*})^n}{n!}\right.\\
  +  A^*C e^{k_2^{'*}m'_2+k_3^{'*}m'_3+k_4^{'*}m'_4}\frac{(k_1^{'*}m'_1)^n}{n!} + AD^* e^{k'_2n_2^{'*}+k'_3n_3^{'*}+k'_4n_4^{'*}}\frac{(k'_1n_1^{'*})^n}{n!} + A^*D e^{k_2^{'*}n'_2+k_3^{'*}n'_3+k_4^{'*}n'_4}\frac{(k_1^{'*}n'_1)^n}{n!}\\
  + BC^* e^{l'_2m_2^{'*}+l'_3m_3^{'*}+l'_4m_4^{'*}}\frac{(l'_1m_1^{'*})^n}{n!} + B^*C e^{l_2^{'*}m'_2+l_3^{'*}m'_3+l_4^{'*}m'_4}\frac{(l_1^{'*}m'_1)^n}{n!} + BD^* e^{l'_2n_2^{'*}+l'_3n_3^{'*}+l'_4n_4^{'*}}\frac{(l'_1n_1^{'*})^n}{n!}\\
  \left.\left.+ B^*D e^{l_2^{'*}n'_2+l_3^{'*}n'_3+l_4^{'*}n'_4}\frac{(l_1^{'*}n'_1)^n}{n!} + CD^* e^{m'_2n_2^{'*}+m'_3n_3^{'*}+m'_4n_4^{'*}}\frac{(m'_1n_1^{'*})^n}{n!}+ C^*D e^{m_2^{'*}n'_2+m_3^{'*}n'_3+m_4^{'*}n'_4}\frac{(m_1^{'*}n'_1)^n}{n!} \right)\right).   \label{B6}
\end{split}
\end{equation}
The probability of getting an even and odd number of photons, i.e., $P(\pm)$, form Eq. \eqref{B6} is written as
    \begin{equation}
\begin{split}
  P(\pm)=|N_j|^2\left(|A|^2\left(\frac{1\pm e^{-2G-W}}{2}\right)\right.+(|B|^2+|D|^2)\left(\frac{1\pm e^{-2G}}{2}\right)+|C|^2\left(\frac{1\pm e^{-2G+W}}{2}\right)\\
  +e^{-(|\alpha|^2+|\zeta|^2)}\left((|A||B|+|A||D|)\left(e^{|\zeta|^2}\cos(q+p)\right.\right.\left.\pm e^{S_1}\cos\left(q-p-W\right)\right)\\
  +(|B||C|+|C||D|)\left(e^{|\zeta|^2}\cos(q+p)\pm e^{S_2}\cos\left(q-p +W\right)\right)+|B||D|\left(e^{|\alpha|^2-|\zeta|^2}\cos(j\pi)\pm e^{S_1-T_1}\cos(j\pi-2W)\right)\\
   \left.\left.+|A||C|\left(e^{|\alpha|^2-|\zeta|^2}\pm e^{S_1-T_1}\right)\cos(j\pi)\right)\right)\label{eqB7}  
\end{split}
\end{equation}
Where,
\begin{equation}
\begin{split}
    G=|\alpha|^2t^2\sin^2(\frac{\phi}{2}) + |\zeta|^2t^2\cos^2(\frac{\phi}{2}),
    T_1=O|\alpha|^2-W,~T_2=O|\alpha|^2 + W,~S_1=U|\zeta|^2 - W,\\
    S_2=U|\zeta|^2 + W,
    U=(-t^2\cos(\phi) + r^2),~W=t^2|\alpha||\zeta|\sin(\phi), O=(t^2\cos(\phi) + r^2).
\end{split}    
\end{equation}
\end{widetext}

\bibliography{apssamp}

\begin{thebibliography}{58}%
\makeatletter
\providecommand \@ifxundefined [1]{%
 \@ifx{#1\undefined}
}%
\providecommand \@ifnum [1]{%
 \ifnum #1\expandafter \@firstoftwo
 \else \expandafter \@secondoftwo
 \fi
}%
\providecommand \@ifx [1]{%
 \ifx #1\expandafter \@firstoftwo
 \else \expandafter \@secondoftwo
 \fi
}%
\providecommand \natexlab [1]{#1}%
\providecommand \enquote  [1]{``#1''}%
\providecommand \bibnamefont  [1]{#1}%
\providecommand \bibfnamefont [1]{#1}%
\providecommand \citenamefont [1]{#1}%
\providecommand \href@noop [0]{\@secondoftwo}%
\providecommand \href [0]{\begingroup \@sanitize@url \@href}%
\providecommand \@href[1]{\@@startlink{#1}\@@href}%
\providecommand \@@href[1]{\endgroup#1\@@endlink}%
\providecommand \@sanitize@url [0]{\catcode `\\12\catcode `\$12\catcode `\&12\catcode `\#12\catcode `\^12\catcode `\_12\catcode `\%12\relax}%
\providecommand \@@startlink[1]{}%
\providecommand \@@endlink[0]{}%
\providecommand \url  [0]{\begingroup\@sanitize@url \@url }%
\providecommand \@url [1]{\endgroup\@href {#1}{\urlprefix }}%
\providecommand \urlprefix  [0]{URL }%
\providecommand \Eprint [0]{\href }%
\providecommand \doibase [0]{https://doi.org/}%
\providecommand \selectlanguage [0]{\@gobble}%
\providecommand \bibinfo  [0]{\@secondoftwo}%
\providecommand \bibfield  [0]{\@secondoftwo}%
\providecommand \translation [1]{[#1]}%
\providecommand \BibitemOpen [0]{}%
\providecommand \bibitemStop [0]{}%
\providecommand \bibitemNoStop [0]{.\EOS\space}%
\providecommand \EOS [0]{\spacefactor3000\relax}%
\providecommand \BibitemShut  [1]{\csname bibitem#1\endcsname}%
\let\auto@bib@innerbib\@empty
\bibitem [{\citenamefont {Mishra}\ \emph {et~al.}(2021)\citenamefont {Mishra}, \citenamefont {Prakash},\ and\ \citenamefont {Jha}}]{mishra2021ququats}%
  \BibitemOpen
  \bibfield  {author} {\bibinfo {author} {\bibfnamefont {M.~K.}\ \bibnamefont {Mishra}}, \bibinfo {author} {\bibfnamefont {H.}~\bibnamefont {Prakash}},\ and\ \bibinfo {author} {\bibfnamefont {V.~B.}\ \bibnamefont {Jha}},\ }\bibfield  {title} {\bibinfo {title} {Ququats as superposition of coherent states and their application in quantum information processing},\ }\href@noop {} {\bibfield  {journal} {\bibinfo  {journal} {International Journal of Quantum Information}\ }\textbf {\bibinfo {volume} {19}},\ \bibinfo {pages} {2150013} (\bibinfo {year} {2021})}\BibitemShut {NoStop}%
\bibitem [{ref(1943)}]{ref1}%
  \BibitemOpen
  \bibfield  {title} {\bibinfo {title} {R{ADIO DETECTION AND RANGING}},\ }\href {https://doi.org/10.1038/152391b0} {\bibfield  {journal} {\bibinfo  {journal} {Nature}\ }\textbf {\bibinfo {volume} {152}},\ \bibinfo {pages} {391} (\bibinfo {year} {1943})}\BibitemShut {NoStop}%
\bibitem [{\citenamefont {Collis}(1970)}]{Collis:70}%
  \BibitemOpen
  \bibfield  {author} {\bibinfo {author} {\bibfnamefont {R.~T.~H.}\ \bibnamefont {Collis}},\ }\bibfield  {title} {\bibinfo {title} {Lidar},\ }\href {https://doi.org/10.1364/AO.9.001782} {\bibfield  {journal} {\bibinfo  {journal} {Appl. Opt.}\ }\textbf {\bibinfo {volume} {9}},\ \bibinfo {pages} {1782} (\bibinfo {year} {1970})}\BibitemShut {NoStop}%
\bibitem [{\citenamefont {Kitchen}\ and\ \citenamefont {Jackson}(1993)}]{kitchen1993weather}%
  \BibitemOpen
  \bibfield  {author} {\bibinfo {author} {\bibfnamefont {M.}~\bibnamefont {Kitchen}}\ and\ \bibinfo {author} {\bibfnamefont {P.}~\bibnamefont {Jackson}},\ }\bibfield  {title} {\bibinfo {title} {Weather radar performance at long range simulated and observed},\ }\href@noop {} {\bibfield  {journal} {\bibinfo  {journal} {Journal of Applied Meteorology and Climatology}\ }\textbf {\bibinfo {volume} {32}},\ \bibinfo {pages} {975} (\bibinfo {year} {1993})}\BibitemShut {NoStop}%
\bibitem [{\citenamefont {Lin}\ \emph {et~al.}(2019)\citenamefont {Lin}, \citenamefont {Cheng}, \citenamefont {Zhou}, \citenamefont {Ravi}, \citenamefont {Hasheminasab}, \citenamefont {Flatt}, \citenamefont {Troy},\ and\ \citenamefont {Habib}}]{lin2019evaluation}%
  \BibitemOpen
  \bibfield  {author} {\bibinfo {author} {\bibfnamefont {Y.-C.}\ \bibnamefont {Lin}}, \bibinfo {author} {\bibfnamefont {Y.-T.}\ \bibnamefont {Cheng}}, \bibinfo {author} {\bibfnamefont {T.}~\bibnamefont {Zhou}}, \bibinfo {author} {\bibfnamefont {R.}~\bibnamefont {Ravi}}, \bibinfo {author} {\bibfnamefont {S.~M.}\ \bibnamefont {Hasheminasab}}, \bibinfo {author} {\bibfnamefont {J.~E.}\ \bibnamefont {Flatt}}, \bibinfo {author} {\bibfnamefont {C.}~\bibnamefont {Troy}},\ and\ \bibinfo {author} {\bibfnamefont {A.}~\bibnamefont {Habib}},\ }\bibfield  {title} {\bibinfo {title} {Evaluation of uav lidar for mapping coastal environments},\ }\href@noop {} {\bibfield  {journal} {\bibinfo  {journal} {Remote Sensing}\ }\textbf {\bibinfo {volume} {11}},\ \bibinfo {pages} {2893} (\bibinfo {year} {2019})}\BibitemShut {NoStop}%
\bibitem [{\citenamefont {Slepyan}\ \emph {et~al.}(2021)\citenamefont {Slepyan}, \citenamefont {Vlasenko}, \citenamefont {Mogilevtsev},\ and\ \citenamefont {Boag}}]{slepyan2021quantum}%
  \BibitemOpen
  \bibfield  {author} {\bibinfo {author} {\bibfnamefont {G.}~\bibnamefont {Slepyan}}, \bibinfo {author} {\bibfnamefont {S.}~\bibnamefont {Vlasenko}}, \bibinfo {author} {\bibfnamefont {D.}~\bibnamefont {Mogilevtsev}},\ and\ \bibinfo {author} {\bibfnamefont {A.}~\bibnamefont {Boag}},\ }\bibfield  {title} {\bibinfo {title} {Quantum radars and lidars: Concepts, realizations, and perspectives},\ }\href@noop {} {\bibfield  {journal} {\bibinfo  {journal} {IEEE Antennas and Propagation Magazine}\ }\textbf {\bibinfo {volume} {64}},\ \bibinfo {pages} {16} (\bibinfo {year} {2021})}\BibitemShut {NoStop}%
\bibitem [{\citenamefont {Wallace}\ \emph {et~al.}(2012)\citenamefont {Wallace}, \citenamefont {Lucieer},\ and\ \citenamefont {Watson}}]{wallace2012assessing}%
  \BibitemOpen
  \bibfield  {author} {\bibinfo {author} {\bibfnamefont {L.}~\bibnamefont {Wallace}}, \bibinfo {author} {\bibfnamefont {A.}~\bibnamefont {Lucieer}},\ and\ \bibinfo {author} {\bibfnamefont {C.}~\bibnamefont {Watson}},\ }\bibfield  {title} {\bibinfo {title} {Assessing the feasibility of uav-based lidar for high resolution forest change detection},\ }\href@noop {} {\bibfield  {journal} {\bibinfo  {journal} {The International Archives of the Photogrammetry, Remote Sensing and Spatial Information Sciences}\ }\textbf {\bibinfo {volume} {39}},\ \bibinfo {pages} {499} (\bibinfo {year} {2012})}\BibitemShut {NoStop}%
\bibitem [{\citenamefont {Blakey}\ \emph {et~al.}(2022)\citenamefont {Blakey}, \citenamefont {Liu}, \citenamefont {Papangelakis}, \citenamefont {Zhang}, \citenamefont {L{\'e}ger}, \citenamefont {Iu},\ and\ \citenamefont {Helmy}}]{blakey2022quantum}%
  \BibitemOpen
  \bibfield  {author} {\bibinfo {author} {\bibfnamefont {P.~S.}\ \bibnamefont {Blakey}}, \bibinfo {author} {\bibfnamefont {H.}~\bibnamefont {Liu}}, \bibinfo {author} {\bibfnamefont {G.}~\bibnamefont {Papangelakis}}, \bibinfo {author} {\bibfnamefont {Y.}~\bibnamefont {Zhang}}, \bibinfo {author} {\bibfnamefont {Z.~M.}\ \bibnamefont {L{\'e}ger}}, \bibinfo {author} {\bibfnamefont {M.~L.}\ \bibnamefont {Iu}},\ and\ \bibinfo {author} {\bibfnamefont {A.~S.}\ \bibnamefont {Helmy}},\ }\bibfield  {title} {\bibinfo {title} {Quantum and non-local effects offer over 40 db noise resilience advantage towards quantum lidar},\ }\href@noop {} {\bibfield  {journal} {\bibinfo  {journal} {Nature communications}\ }\textbf {\bibinfo {volume} {13}},\ \bibinfo {pages} {5633} (\bibinfo {year} {2022})}\BibitemShut {NoStop}%
\bibitem [{\citenamefont {Liu}\ \emph {et~al.}(2023)\citenamefont {Liu}, \citenamefont {Qin}, \citenamefont {Papangelakis}, \citenamefont {Iu},\ and\ \citenamefont {Helmy}}]{liu2023compact}%
  \BibitemOpen
  \bibfield  {author} {\bibinfo {author} {\bibfnamefont {H.}~\bibnamefont {Liu}}, \bibinfo {author} {\bibfnamefont {C.}~\bibnamefont {Qin}}, \bibinfo {author} {\bibfnamefont {G.}~\bibnamefont {Papangelakis}}, \bibinfo {author} {\bibfnamefont {M.~L.}\ \bibnamefont {Iu}},\ and\ \bibinfo {author} {\bibfnamefont {A.~S.}\ \bibnamefont {Helmy}},\ }\bibfield  {title} {\bibinfo {title} {Compact all-fiber quantum-inspired {LiDAR} with over 100 db noise rejection and single photon sensitivity},\ }\href@noop {} {\bibfield  {journal} {\bibinfo  {journal} {Nature Communications}\ }\textbf {\bibinfo {volume} {14}},\ \bibinfo {pages} {5344} (\bibinfo {year} {2023})}\BibitemShut {NoStop}%
\bibitem [{\citenamefont {Tsai}\ and\ \citenamefont {Liu}(2020)}]{tsai2020anti}%
  \BibitemOpen
  \bibfield  {author} {\bibinfo {author} {\bibfnamefont {C.-M.}\ \bibnamefont {Tsai}}\ and\ \bibinfo {author} {\bibfnamefont {Y.-C.}\ \bibnamefont {Liu}},\ }\bibfield  {title} {\bibinfo {title} {Anti-interference single-photon lidar using stochastic pulse position modulation},\ }\href@noop {} {\bibfield  {journal} {\bibinfo  {journal} {Optics Letters}\ }\textbf {\bibinfo {volume} {45}},\ \bibinfo {pages} {439} (\bibinfo {year} {2020})}\BibitemShut {NoStop}%
\bibitem [{\citenamefont {Laurenzis}\ \emph {et~al.}(2017)\citenamefont {Laurenzis}, \citenamefont {Hengy}, \citenamefont {Hommes}, \citenamefont {Kloeppel}, \citenamefont {Shoykhetbrod}, \citenamefont {Geibig}, \citenamefont {Johannes}, \citenamefont {Naz},\ and\ \citenamefont {Christnacher}}]{laurenzis2017multi}%
  \BibitemOpen
  \bibfield  {author} {\bibinfo {author} {\bibfnamefont {M.}~\bibnamefont {Laurenzis}}, \bibinfo {author} {\bibfnamefont {S.}~\bibnamefont {Hengy}}, \bibinfo {author} {\bibfnamefont {A.}~\bibnamefont {Hommes}}, \bibinfo {author} {\bibfnamefont {F.}~\bibnamefont {Kloeppel}}, \bibinfo {author} {\bibfnamefont {A.}~\bibnamefont {Shoykhetbrod}}, \bibinfo {author} {\bibfnamefont {T.}~\bibnamefont {Geibig}}, \bibinfo {author} {\bibfnamefont {W.}~\bibnamefont {Johannes}}, \bibinfo {author} {\bibfnamefont {P.}~\bibnamefont {Naz}},\ and\ \bibinfo {author} {\bibfnamefont {F.}~\bibnamefont {Christnacher}},\ }\bibfield  {title} {\bibinfo {title} {Multi-sensor field trials for detection and tracking of multiple small unmanned aerial vehicles flying at low altitude},\ }in\ \href@noop {} {\emph {\bibinfo {booktitle} {Signal Processing, Sensor/Information Fusion, and Target Recognition XXVI}}},\ Vol.\ \bibinfo {volume} {10200}\ (\bibinfo {organization} {SPIE},\ \bibinfo {year} {2017})\ pp.\ \bibinfo {pages}
  {384--396}\BibitemShut {NoStop}%
\bibitem [{\citenamefont {Kim}\ \emph {et~al.}(2021)\citenamefont {Kim}, \citenamefont {Martins}, \citenamefont {Jang}, \citenamefont {Badloe}, \citenamefont {Khadir}, \citenamefont {Jung}, \citenamefont {Kim}, \citenamefont {Kim}, \citenamefont {Genevet},\ and\ \citenamefont {Rho}}]{kim2021nanophotonics}%
  \BibitemOpen
  \bibfield  {author} {\bibinfo {author} {\bibfnamefont {I.}~\bibnamefont {Kim}}, \bibinfo {author} {\bibfnamefont {R.~J.}\ \bibnamefont {Martins}}, \bibinfo {author} {\bibfnamefont {J.}~\bibnamefont {Jang}}, \bibinfo {author} {\bibfnamefont {T.}~\bibnamefont {Badloe}}, \bibinfo {author} {\bibfnamefont {S.}~\bibnamefont {Khadir}}, \bibinfo {author} {\bibfnamefont {H.-Y.}\ \bibnamefont {Jung}}, \bibinfo {author} {\bibfnamefont {H.}~\bibnamefont {Kim}}, \bibinfo {author} {\bibfnamefont {J.}~\bibnamefont {Kim}}, \bibinfo {author} {\bibfnamefont {P.}~\bibnamefont {Genevet}},\ and\ \bibinfo {author} {\bibfnamefont {J.}~\bibnamefont {Rho}},\ }\bibfield  {title} {\bibinfo {title} {Nanophotonics for light detection and ranging technology},\ }\href@noop {} {\bibfield  {journal} {\bibinfo  {journal} {Nature nanotechnology}\ }\textbf {\bibinfo {volume} {16}},\ \bibinfo {pages} {508} (\bibinfo {year} {2021})}\BibitemShut {NoStop}%
\bibitem [{\citenamefont {Helstrom}(1976)}]{1976Helstrom}%
  \BibitemOpen
  \bibfield  {author} {\bibinfo {author} {\bibfnamefont {C.~W.}\ \bibnamefont {Helstrom}},\ }\href {https://books.google.co.in/books?id=Ne3iT\_QLcsMC} {\emph {\bibinfo {title} {Quantum {D}etection and {E}stimation {T}heory}}}\ (\bibinfo  {publisher} {Academic Press, San Diego, CA},\ \bibinfo {year} {1976})\BibitemShut {NoStop}%
\bibitem [{\citenamefont {Giovannetti}\ \emph {et~al.}(2006)\citenamefont {Giovannetti}, \citenamefont {Lloyd},\ and\ \citenamefont {Maccone}}]{giovannetti2006quantum}%
  \BibitemOpen
  \bibfield  {author} {\bibinfo {author} {\bibfnamefont {V.}~\bibnamefont {Giovannetti}}, \bibinfo {author} {\bibfnamefont {S.}~\bibnamefont {Lloyd}},\ and\ \bibinfo {author} {\bibfnamefont {L.}~\bibnamefont {Maccone}},\ }\bibfield  {title} {\bibinfo {title} {Quantum metrology},\ }\href@noop {} {\bibfield  {journal} {\bibinfo  {journal} {Physical review letters}\ }\textbf {\bibinfo {volume} {96}},\ \bibinfo {pages} {010401} (\bibinfo {year} {2006})}\BibitemShut {NoStop}%
\bibitem [{\citenamefont {Dowling}(2009)}]{dowling2009quantum}%
  \BibitemOpen
  \bibfield  {author} {\bibinfo {author} {\bibfnamefont {J.}~\bibnamefont {Dowling}},\ }\bibfield  {title} {\bibinfo {title} {Quantum lidar-remote sensing at the ultimate limit},\ }in\ \href@noop {} {\emph {\bibinfo {booktitle} {Louisiana State University Baton Rouge, AIR FORCE RE-SEARCH LABORATORY INFORMATION DIRECTORATE ROME RESEARCH SITE ROME}}}\ (\bibinfo  {publisher} {Tech. Rep.},\ \bibinfo {year} {2009})\BibitemShut {NoStop}%
\bibitem [{\citenamefont {Dowling}(2008{\natexlab{a}})}]{doi:10.1080/00107510802091298}%
  \BibitemOpen
  \bibfield  {author} {\bibinfo {author} {\bibfnamefont {J.~P.}\ \bibnamefont {Dowling}},\ }\bibfield  {title} {\bibinfo {title} {Quantum optical metrology – the lowdown on high-{N00N} states},\ }\href {https://doi.org/10.1080/00107510802091298} {\bibfield  {journal} {\bibinfo  {journal} {Contemporary Physics}\ }\textbf {\bibinfo {volume} {49}},\ \bibinfo {pages} {125} (\bibinfo {year} {2008}{\natexlab{a}})}\BibitemShut {NoStop}%
\bibitem [{\citenamefont {Born}\ and\ \citenamefont {Wolf}(2013)}]{born2013principles}%
  \BibitemOpen
  \bibfield  {author} {\bibinfo {author} {\bibfnamefont {M.}~\bibnamefont {Born}}\ and\ \bibinfo {author} {\bibfnamefont {E.}~\bibnamefont {Wolf}},\ }\href@noop {} {\emph {\bibinfo {title} {Principles of optics: electromagnetic theory of propagation, interference and diffraction of light}}}\ (\bibinfo  {publisher} {Elsevier},\ \bibinfo {year} {2013})\BibitemShut {NoStop}%
\bibitem [{\citenamefont {Boto}\ \emph {et~al.}(2000)\citenamefont {Boto}, \citenamefont {Kok}, \citenamefont {Abrams}, \citenamefont {Braunstein}, \citenamefont {Williams},\ and\ \citenamefont {Dowling}}]{boto2000quantum}%
  \BibitemOpen
  \bibfield  {author} {\bibinfo {author} {\bibfnamefont {A.~N.}\ \bibnamefont {Boto}}, \bibinfo {author} {\bibfnamefont {P.}~\bibnamefont {Kok}}, \bibinfo {author} {\bibfnamefont {D.~S.}\ \bibnamefont {Abrams}}, \bibinfo {author} {\bibfnamefont {S.~L.}\ \bibnamefont {Braunstein}}, \bibinfo {author} {\bibfnamefont {C.~P.}\ \bibnamefont {Williams}},\ and\ \bibinfo {author} {\bibfnamefont {J.~P.}\ \bibnamefont {Dowling}},\ }\bibfield  {title} {\bibinfo {title} {Quantum interferometric optical lithography: exploiting entanglement to beat the diffraction limit},\ }\href@noop {} {\bibfield  {journal} {\bibinfo  {journal} {Physical Review Letters}\ }\textbf {\bibinfo {volume} {85}},\ \bibinfo {pages} {2733} (\bibinfo {year} {2000})}\BibitemShut {NoStop}%
\bibitem [{\citenamefont {Mitchell}\ \emph {et~al.}(2004)\citenamefont {Mitchell}, \citenamefont {Lundeen},\ and\ \citenamefont {Steinberg}}]{mitchell2004super}%
  \BibitemOpen
  \bibfield  {author} {\bibinfo {author} {\bibfnamefont {M.~W.}\ \bibnamefont {Mitchell}}, \bibinfo {author} {\bibfnamefont {J.~S.}\ \bibnamefont {Lundeen}},\ and\ \bibinfo {author} {\bibfnamefont {A.~M.}\ \bibnamefont {Steinberg}},\ }\bibfield  {title} {\bibinfo {title} {Super-resolving phase measurements with a multiphoton entangled state},\ }\href@noop {} {\bibfield  {journal} {\bibinfo  {journal} {Nature}\ }\textbf {\bibinfo {volume} {429}},\ \bibinfo {pages} {161} (\bibinfo {year} {2004})}\BibitemShut {NoStop}%
\bibitem [{\citenamefont {Resch}\ \emph {et~al.}(2007)\citenamefont {Resch}, \citenamefont {Pregnell}, \citenamefont {Prevedel}, \citenamefont {Gilchrist}, \citenamefont {Pryde}, \citenamefont {O'Brien},\ and\ \citenamefont {White}}]{PhysRevLett.98.223601}%
  \BibitemOpen
  \bibfield  {author} {\bibinfo {author} {\bibfnamefont {K.~J.}\ \bibnamefont {Resch}}, \bibinfo {author} {\bibfnamefont {K.~L.}\ \bibnamefont {Pregnell}}, \bibinfo {author} {\bibfnamefont {R.}~\bibnamefont {Prevedel}}, \bibinfo {author} {\bibfnamefont {A.}~\bibnamefont {Gilchrist}}, \bibinfo {author} {\bibfnamefont {G.~J.}\ \bibnamefont {Pryde}}, \bibinfo {author} {\bibfnamefont {J.~L.}\ \bibnamefont {O'Brien}},\ and\ \bibinfo {author} {\bibfnamefont {A.~G.}\ \bibnamefont {White}},\ }\bibfield  {title} {\bibinfo {title} {Time-{R}eversal and {S}uper-{R}esolving {P}hase {M}easurements},\ }\href {https://doi.org/10.1103/PhysRevLett.98.223601} {\bibfield  {journal} {\bibinfo  {journal} {Phys. Rev. Lett.}\ }\textbf {\bibinfo {volume} {98}},\ \bibinfo {pages} {223601} (\bibinfo {year} {2007})}\BibitemShut {NoStop}%
\bibitem [{\citenamefont {Gao}\ \emph {et~al.}(2010)\citenamefont {Gao}, \citenamefont {Anisimov}, \citenamefont {Wildfeuer}, \citenamefont {Luine}, \citenamefont {Lee},\ and\ \citenamefont {Dowling}}]{Gao:10}%
  \BibitemOpen
  \bibfield  {author} {\bibinfo {author} {\bibfnamefont {Y.}~\bibnamefont {Gao}}, \bibinfo {author} {\bibfnamefont {P.~M.}\ \bibnamefont {Anisimov}}, \bibinfo {author} {\bibfnamefont {C.~F.}\ \bibnamefont {Wildfeuer}}, \bibinfo {author} {\bibfnamefont {J.}~\bibnamefont {Luine}}, \bibinfo {author} {\bibfnamefont {H.}~\bibnamefont {Lee}},\ and\ \bibinfo {author} {\bibfnamefont {J.~P.}\ \bibnamefont {Dowling}},\ }\bibfield  {title} {\bibinfo {title} {Super-resolution at the shot-noise limit with coherent states and photon-number-resolving detectors},\ }\href {https://doi.org/10.1364/JOSAB.27.00A170} {\bibfield  {journal} {\bibinfo  {journal} {J. Opt. Soc. Am. B}\ }\textbf {\bibinfo {volume} {27}},\ \bibinfo {pages} {A170} (\bibinfo {year} {2010})}\BibitemShut {NoStop}%
\bibitem [{\citenamefont {Distante}\ \emph {et~al.}(2013)\citenamefont {Distante}, \citenamefont {Je\ifmmode~\check{z}\else \v{z}\fi{}ek},\ and\ \citenamefont {Andersen}}]{PhysRevLett.111.033603}%
  \BibitemOpen
  \bibfield  {author} {\bibinfo {author} {\bibfnamefont {E.}~\bibnamefont {Distante}}, \bibinfo {author} {\bibfnamefont {M.}~\bibnamefont {Je\ifmmode~\check{z}\else \v{z}\fi{}ek}},\ and\ \bibinfo {author} {\bibfnamefont {U.~L.}\ \bibnamefont {Andersen}},\ }\bibfield  {title} {\bibinfo {title} {Deterministic superresolution with coherent states at the shot noise limit},\ }\href {https://doi.org/10.1103/PhysRevLett.111.033603} {\bibfield  {journal} {\bibinfo  {journal} {Phys. Rev. Lett.}\ }\textbf {\bibinfo {volume} {111}},\ \bibinfo {pages} {033603} (\bibinfo {year} {2013})}\BibitemShut {NoStop}%
\bibitem [{\citenamefont {Caves}(1981)}]{PhysRevD.23.1693}%
  \BibitemOpen
  \bibfield  {author} {\bibinfo {author} {\bibfnamefont {C.~M.}\ \bibnamefont {Caves}},\ }\bibfield  {title} {\bibinfo {title} {Quantum-mechanical noise in an interferometer},\ }\href {https://doi.org/10.1103/PhysRevD.23.1693} {\bibfield  {journal} {\bibinfo  {journal} {Phys. Rev. D}\ }\textbf {\bibinfo {volume} {23}},\ \bibinfo {pages} {1693} (\bibinfo {year} {1981})}\BibitemShut {NoStop}%
\bibitem [{\citenamefont {Xiao}\ \emph {et~al.}(1987)\citenamefont {Xiao}, \citenamefont {Wu},\ and\ \citenamefont {Kimble}}]{xiao1987precision}%
  \BibitemOpen
  \bibfield  {author} {\bibinfo {author} {\bibfnamefont {M.}~\bibnamefont {Xiao}}, \bibinfo {author} {\bibfnamefont {L.-A.}\ \bibnamefont {Wu}},\ and\ \bibinfo {author} {\bibfnamefont {H.~J.}\ \bibnamefont {Kimble}},\ }\bibfield  {title} {\bibinfo {title} {Precision measurement beyond the shot-noise limit},\ }\href@noop {} {\bibfield  {journal} {\bibinfo  {journal} {Physical review letters}\ }\textbf {\bibinfo {volume} {59}},\ \bibinfo {pages} {278} (\bibinfo {year} {1987})}\BibitemShut {NoStop}%
\bibitem [{\citenamefont {Shukla}\ \emph {et~al.}(2024)\citenamefont {Shukla}, \citenamefont {Yadav}, \citenamefont {Sharma}, \citenamefont {Kumar},\ and\ \citenamefont {Mishra}}]{shukla2024quantum}%
  \BibitemOpen
  \bibfield  {author} {\bibinfo {author} {\bibfnamefont {G.}~\bibnamefont {Shukla}}, \bibinfo {author} {\bibfnamefont {D.}~\bibnamefont {Yadav}}, \bibinfo {author} {\bibfnamefont {P.}~\bibnamefont {Sharma}}, \bibinfo {author} {\bibfnamefont {A.}~\bibnamefont {Kumar}},\ and\ \bibinfo {author} {\bibfnamefont {D.~K.}\ \bibnamefont {Mishra}},\ }\bibfield  {title} {\bibinfo {title} {Quantum sub-phase sensitivity of a mach--zehnder interferometer with the superposition of schr{\"o}dinger’s cat-like state with vacuum state as an input under product detection scheme},\ }\href@noop {} {\bibfield  {journal} {\bibinfo  {journal} {Physics Open}\ }\textbf {\bibinfo {volume} {18}},\ \bibinfo {pages} {100200} (\bibinfo {year} {2024})}\BibitemShut {NoStop}%
\bibitem [{\citenamefont {Shukla}\ \emph {et~al.}(2023)\citenamefont {Shukla}, \citenamefont {Mishra}, \citenamefont {Pandey}, \citenamefont {Kumar}, \citenamefont {Pandey},\ and\ \citenamefont {Mishra}}]{shukla2023improvement}%
  \BibitemOpen
  \bibfield  {author} {\bibinfo {author} {\bibfnamefont {G.}~\bibnamefont {Shukla}}, \bibinfo {author} {\bibfnamefont {K.~M.}\ \bibnamefont {Mishra}}, \bibinfo {author} {\bibfnamefont {A.~K.}\ \bibnamefont {Pandey}}, \bibinfo {author} {\bibfnamefont {T.}~\bibnamefont {Kumar}}, \bibinfo {author} {\bibfnamefont {H.}~\bibnamefont {Pandey}},\ and\ \bibinfo {author} {\bibfnamefont {D.~K.}\ \bibnamefont {Mishra}},\ }\bibfield  {title} {\bibinfo {title} {Improvement in phase-sensitivity of a mach--zehnder interferometer with the superposition of schr{\"o}dinger’s cat-like state with vacuum state as an input under parity measurement},\ }\href@noop {} {\bibfield  {journal} {\bibinfo  {journal} {Optical and Quantum Electronics}\ }\textbf {\bibinfo {volume} {55}},\ \bibinfo {pages} {460} (\bibinfo {year} {2023})}\BibitemShut {NoStop}%
\bibitem [{\citenamefont {Dowling}(2008{\natexlab{b}})}]{dowling2008quantum}%
  \BibitemOpen
  \bibfield  {author} {\bibinfo {author} {\bibfnamefont {J.~P.}\ \bibnamefont {Dowling}},\ }\bibfield  {title} {\bibinfo {title} {Quantum optical metrology--the lowdown on high-n00n states},\ }\href@noop {} {\bibfield  {journal} {\bibinfo  {journal} {Contemporary physics}\ }\textbf {\bibinfo {volume} {49}},\ \bibinfo {pages} {125} (\bibinfo {year} {2008}{\natexlab{b}})}\BibitemShut {NoStop}%
\bibitem [{\citenamefont {Shukla}\ \emph {et~al.}(2022)\citenamefont {Shukla}, \citenamefont {Mishra}, \citenamefont {Yadav}, \citenamefont {Pandey},\ and\ \citenamefont {Mishra}}]{Shukla:22}%
  \BibitemOpen
  \bibfield  {author} {\bibinfo {author} {\bibfnamefont {G.}~\bibnamefont {Shukla}}, \bibinfo {author} {\bibfnamefont {K.~K.}\ \bibnamefont {Mishra}}, \bibinfo {author} {\bibfnamefont {D.}~\bibnamefont {Yadav}}, \bibinfo {author} {\bibfnamefont {R.~K.}\ \bibnamefont {Pandey}},\ and\ \bibinfo {author} {\bibfnamefont {D.~K.}\ \bibnamefont {Mishra}},\ }\bibfield  {title} {\bibinfo {title} {Quantum-enhanced super-sensitivity of a {M}ach--{Z}ehnder interferometer with superposition of {S}chr\"{o}dinger's cat-like state and {F}ock state as inputs using a two-channel detection},\ }\href {https://doi.org/10.1364/JOSAB.434967} {\bibfield  {journal} {\bibinfo  {journal} {J. Opt. Soc. Am. B}\ }\textbf {\bibinfo {volume} {39}},\ \bibinfo {pages} {59} (\bibinfo {year} {2022})}\BibitemShut {NoStop}%
\bibitem [{\citenamefont {Giovannetti}\ \emph {et~al.}(2011)\citenamefont {Giovannetti}, \citenamefont {Lloyd},\ and\ \citenamefont {Maccone}}]{giovannetti2011advances}%
  \BibitemOpen
  \bibfield  {author} {\bibinfo {author} {\bibfnamefont {V.}~\bibnamefont {Giovannetti}}, \bibinfo {author} {\bibfnamefont {S.}~\bibnamefont {Lloyd}},\ and\ \bibinfo {author} {\bibfnamefont {L.}~\bibnamefont {Maccone}},\ }\bibfield  {title} {\bibinfo {title} {Advances in quantum metrology},\ }\href@noop {} {\bibfield  {journal} {\bibinfo  {journal} {Nature photonics}\ }\textbf {\bibinfo {volume} {5}},\ \bibinfo {pages} {222} (\bibinfo {year} {2011})}\BibitemShut {NoStop}%
\bibitem [{\citenamefont {Shukla}\ \emph {et~al.}(2021)\citenamefont {Shukla}, \citenamefont {Salykina}, \citenamefont {Frascella}, \citenamefont {Mishra}, \citenamefont {Chekhova},\ and\ \citenamefont {Khalili}}]{Shukla:21}%
  \BibitemOpen
  \bibfield  {author} {\bibinfo {author} {\bibfnamefont {G.}~\bibnamefont {Shukla}}, \bibinfo {author} {\bibfnamefont {D.}~\bibnamefont {Salykina}}, \bibinfo {author} {\bibfnamefont {G.}~\bibnamefont {Frascella}}, \bibinfo {author} {\bibfnamefont {D.~K.}\ \bibnamefont {Mishra}}, \bibinfo {author} {\bibfnamefont {M.~V.}\ \bibnamefont {Chekhova}},\ and\ \bibinfo {author} {\bibfnamefont {F.~Y.}\ \bibnamefont {Khalili}},\ }\bibfield  {title} {\bibinfo {title} {Broadening the high sensitivity range of squeezing-assisted interferometers by means of two-channel detection},\ }\href {https://doi.org/10.1364/OE.413391} {\bibfield  {journal} {\bibinfo  {journal} {Opt. Express}\ }\textbf {\bibinfo {volume} {29}},\ \bibinfo {pages} {95} (\bibinfo {year} {2021})}\BibitemShut {NoStop}%
\bibitem [{\citenamefont {Huver}\ \emph {et~al.}(2008)\citenamefont {Huver}, \citenamefont {Wildfeuer},\ and\ \citenamefont {Dowling}}]{PhysRevA.78.063828}%
  \BibitemOpen
  \bibfield  {author} {\bibinfo {author} {\bibfnamefont {S.~D.}\ \bibnamefont {Huver}}, \bibinfo {author} {\bibfnamefont {C.~F.}\ \bibnamefont {Wildfeuer}},\ and\ \bibinfo {author} {\bibfnamefont {J.~P.}\ \bibnamefont {Dowling}},\ }\bibfield  {title} {\bibinfo {title} {Entangled {F}ock states for robust quantum optical metrology, imaging, and sensing},\ }\href {https://doi.org/10.1103/PhysRevA.78.063828} {\bibfield  {journal} {\bibinfo  {journal} {Phys. Rev. A}\ }\textbf {\bibinfo {volume} {78}},\ \bibinfo {pages} {063828} (\bibinfo {year} {2008})}\BibitemShut {NoStop}%
\bibitem [{\citenamefont {Rubin}\ and\ \citenamefont {Kaushik}(2007)}]{PhysRevA.75.053805}%
  \BibitemOpen
  \bibfield  {author} {\bibinfo {author} {\bibfnamefont {M.~A.}\ \bibnamefont {Rubin}}\ and\ \bibinfo {author} {\bibfnamefont {S.}~\bibnamefont {Kaushik}},\ }\bibfield  {title} {\bibinfo {title} {Loss-induced limits to phase measurement precision with maximally entangled states},\ }\href {https://doi.org/10.1103/PhysRevA.75.053805} {\bibfield  {journal} {\bibinfo  {journal} {Phys. Rev. A}\ }\textbf {\bibinfo {volume} {75}},\ \bibinfo {pages} {053805} (\bibinfo {year} {2007})}\BibitemShut {NoStop}%
\bibitem [{\citenamefont {Reichert}\ \emph {et~al.}(2022)\citenamefont {Reichert}, \citenamefont {Di~Candia}, \citenamefont {Win},\ and\ \citenamefont {Sanz}}]{reichert2022quantum}%
  \BibitemOpen
  \bibfield  {author} {\bibinfo {author} {\bibfnamefont {M.}~\bibnamefont {Reichert}}, \bibinfo {author} {\bibfnamefont {R.}~\bibnamefont {Di~Candia}}, \bibinfo {author} {\bibfnamefont {M.~Z.}\ \bibnamefont {Win}},\ and\ \bibinfo {author} {\bibfnamefont {M.}~\bibnamefont {Sanz}},\ }\bibfield  {title} {\bibinfo {title} {Quantum-enhanced {D}oppler lidar},\ }\href@noop {} {\bibfield  {journal} {\bibinfo  {journal} {npj Quantum Information}\ }\textbf {\bibinfo {volume} {8}},\ \bibinfo {pages} {147} (\bibinfo {year} {2022})}\BibitemShut {NoStop}%
\bibitem [{\citenamefont {Wang}\ \emph {et~al.}(2016{\natexlab{a}})\citenamefont {Wang}, \citenamefont {Hao}, \citenamefont {Tang}, \citenamefont {Zhang}, \citenamefont {Yang}, \citenamefont {Yang}, \citenamefont {Xu},\ and\ \citenamefont {Zhao}}]{WANG20163717}%
  \BibitemOpen
  \bibfield  {author} {\bibinfo {author} {\bibfnamefont {Q.}~\bibnamefont {Wang}}, \bibinfo {author} {\bibfnamefont {L.}~\bibnamefont {Hao}}, \bibinfo {author} {\bibfnamefont {H.}~\bibnamefont {Tang}}, \bibinfo {author} {\bibfnamefont {Y.}~\bibnamefont {Zhang}}, \bibinfo {author} {\bibfnamefont {C.}~\bibnamefont {Yang}}, \bibinfo {author} {\bibfnamefont {X.}~\bibnamefont {Yang}}, \bibinfo {author} {\bibfnamefont {L.}~\bibnamefont {Xu}},\ and\ \bibinfo {author} {\bibfnamefont {Y.}~\bibnamefont {Zhao}},\ }\bibfield  {title} {\bibinfo {title} {Super-resolving quantum {LiDAR} with even coherent states sources in the presence of loss and noise},\ }\href {https://doi.org/https://doi.org/10.1016/j.physleta.2016.08.033} {\bibfield  {journal} {\bibinfo  {journal} {Physics Letters A}\ }\textbf {\bibinfo {volume} {380}},\ \bibinfo {pages} {3717} (\bibinfo {year} {2016}{\natexlab{a}})}\BibitemShut {NoStop}%
\bibitem [{\citenamefont {Dodonov}\ \emph {et~al.}(1974)\citenamefont {Dodonov}, \citenamefont {Malkin},\ and\ \citenamefont {Man'ko}}]{DODONOV1974597}%
  \BibitemOpen
  \bibfield  {author} {\bibinfo {author} {\bibfnamefont {V.}~\bibnamefont {Dodonov}}, \bibinfo {author} {\bibfnamefont {I.}~\bibnamefont {Malkin}},\ and\ \bibinfo {author} {\bibfnamefont {V.}~\bibnamefont {Man'ko}},\ }\bibfield  {title} {\bibinfo {title} {Even and odd coherent states and excitations of a singular oscillator},\ }\href {https://doi.org/https://doi.org/10.1016/0031-8914(74)90215-8} {\bibfield  {journal} {\bibinfo  {journal} {Physica}\ }\textbf {\bibinfo {volume} {72}},\ \bibinfo {pages} {597} (\bibinfo {year} {1974})}\BibitemShut {NoStop}%
\bibitem [{\citenamefont {Mishra}\ and\ \citenamefont {Prakash}(2010)}]{mishra2010teleportation}%
  \BibitemOpen
  \bibfield  {author} {\bibinfo {author} {\bibfnamefont {M.~K.}\ \bibnamefont {Mishra}}\ and\ \bibinfo {author} {\bibfnamefont {H.}~\bibnamefont {Prakash}},\ }\bibfield  {title} {\bibinfo {title} {Teleportation of a two-mode entangled coherent state encoded with two-qubit information},\ }\href@noop {} {\bibfield  {journal} {\bibinfo  {journal} {Journal of Physics B: Atomic, Molecular and Optical Physics}\ }\textbf {\bibinfo {volume} {43}},\ \bibinfo {pages} {185501} (\bibinfo {year} {2010})}\BibitemShut {NoStop}%
\bibitem [{\citenamefont {Prakash}\ and\ \citenamefont {Mishra}(2012)}]{prakash2012teleportation}%
  \BibitemOpen
  \bibfield  {author} {\bibinfo {author} {\bibfnamefont {H.}~\bibnamefont {Prakash}}\ and\ \bibinfo {author} {\bibfnamefont {M.~K.}\ \bibnamefont {Mishra}},\ }\bibfield  {title} {\bibinfo {title} {Teleportation of superposed coherent states using nonmaximally entangled resources},\ }\href@noop {} {\bibfield  {journal} {\bibinfo  {journal} {JOSA B}\ }\textbf {\bibinfo {volume} {29}},\ \bibinfo {pages} {2915} (\bibinfo {year} {2012})}\BibitemShut {NoStop}%
\bibitem [{\citenamefont {Mishra}\ \emph {et~al.}(2016)\citenamefont {Mishra}, \citenamefont {Maurya},\ and\ \citenamefont {Prakash}}]{mishra2016quantum}%
  \BibitemOpen
  \bibfield  {author} {\bibinfo {author} {\bibfnamefont {M.~K.}\ \bibnamefont {Mishra}}, \bibinfo {author} {\bibfnamefont {A.~K.}\ \bibnamefont {Maurya}},\ and\ \bibinfo {author} {\bibfnamefont {H.}~\bibnamefont {Prakash}},\ }\bibfield  {title} {\bibinfo {title} {Quantum discord and entanglement of quasi-werner states based on bipartite entangled coherent states},\ }\href@noop {} {\bibfield  {journal} {\bibinfo  {journal} {International Journal of Theoretical Physics}\ }\textbf {\bibinfo {volume} {55}},\ \bibinfo {pages} {2735} (\bibinfo {year} {2016})}\BibitemShut {NoStop}%
\bibitem [{\citenamefont {Maurya}\ \emph {et~al.}(2016)\citenamefont {Maurya}, \citenamefont {Mishra},\ and\ \citenamefont {Prakash}}]{maurya2016two}%
  \BibitemOpen
  \bibfield  {author} {\bibinfo {author} {\bibfnamefont {A.~K.}\ \bibnamefont {Maurya}}, \bibinfo {author} {\bibfnamefont {M.~K.}\ \bibnamefont {Mishra}},\ and\ \bibinfo {author} {\bibfnamefont {H.}~\bibnamefont {Prakash}},\ }\bibfield  {title} {\bibinfo {title} {Two-way quantum communication: Generalization of secure quantum information exchange to quantum network},\ }\href@noop {} {\bibfield  {journal} {\bibinfo  {journal} {Pramana}\ }\textbf {\bibinfo {volume} {86}},\ \bibinfo {pages} {515} (\bibinfo {year} {2016})}\BibitemShut {NoStop}%
\bibitem [{\citenamefont {Mishra}\ \emph {et~al.}(2011)\citenamefont {Mishra}, \citenamefont {Maurya},\ and\ \citenamefont {Prakash}}]{mishra2011two}%
  \BibitemOpen
  \bibfield  {author} {\bibinfo {author} {\bibfnamefont {M.~K.}\ \bibnamefont {Mishra}}, \bibinfo {author} {\bibfnamefont {A.~K.}\ \bibnamefont {Maurya}},\ and\ \bibinfo {author} {\bibfnamefont {H.}~\bibnamefont {Prakash}},\ }\bibfield  {title} {\bibinfo {title} {Two-way quantum communication:‘secure quantum information exchange’},\ }\href@noop {} {\bibfield  {journal} {\bibinfo  {journal} {Journal of Physics B: Atomic, Molecular and Optical Physics}\ }\textbf {\bibinfo {volume} {44}},\ \bibinfo {pages} {115504} (\bibinfo {year} {2011})}\BibitemShut {NoStop}%
\bibitem [{\citenamefont {Mishra}\ and\ \citenamefont {Prakash}(2013)}]{mishra2013bipartite}%
  \BibitemOpen
  \bibfield  {author} {\bibinfo {author} {\bibfnamefont {M.~K.}\ \bibnamefont {Mishra}}\ and\ \bibinfo {author} {\bibfnamefont {H.}~\bibnamefont {Prakash}},\ }\bibfield  {title} {\bibinfo {title} {Bipartite coherent-state quantum key distribution with strong reference pulse},\ }\href@noop {} {\bibfield  {journal} {\bibinfo  {journal} {Quantum information processing}\ }\textbf {\bibinfo {volume} {12}},\ \bibinfo {pages} {907} (\bibinfo {year} {2013})}\BibitemShut {NoStop}%
\bibitem [{\citenamefont {Wigner}(1932)}]{wigner1932quantum}%
  \BibitemOpen
  \bibfield  {author} {\bibinfo {author} {\bibfnamefont {E.}~\bibnamefont {Wigner}},\ }\bibfield  {title} {\bibinfo {title} {On the quantum correction for thermodynamic equilibrium},\ }\href@noop {} {\bibfield  {journal} {\bibinfo  {journal} {Physical review}\ }\textbf {\bibinfo {volume} {40}},\ \bibinfo {pages} {749} (\bibinfo {year} {1932})}\BibitemShut {NoStop}%
\bibitem [{\citenamefont {Hillery}\ \emph {et~al.}(1984)\citenamefont {Hillery}, \citenamefont {O'Connell}, \citenamefont {Scully},\ and\ \citenamefont {Wigner}}]{hillery1984distribution}%
  \BibitemOpen
  \bibfield  {author} {\bibinfo {author} {\bibfnamefont {M.}~\bibnamefont {Hillery}}, \bibinfo {author} {\bibfnamefont {R.~F.}\ \bibnamefont {O'Connell}}, \bibinfo {author} {\bibfnamefont {M.~O.}\ \bibnamefont {Scully}},\ and\ \bibinfo {author} {\bibfnamefont {E.~P.}\ \bibnamefont {Wigner}},\ }\bibfield  {title} {\bibinfo {title} {Distribution functions in physics: Fundamentals},\ }\href@noop {} {\bibfield  {journal} {\bibinfo  {journal} {Physics reports}\ }\textbf {\bibinfo {volume} {106}},\ \bibinfo {pages} {121} (\bibinfo {year} {1984})}\BibitemShut {NoStop}%
\bibitem [{\citenamefont {Loudon}(2000)}]{loudon2000quantum}%
  \BibitemOpen
  \bibfield  {author} {\bibinfo {author} {\bibfnamefont {R.}~\bibnamefont {Loudon}},\ }\href {https://books.google.co.in/books?id=AEkfajgqldoC} {\emph {\bibinfo {title} {The Quantum Theory of Light}}}\ (\bibinfo  {publisher} {OUP Oxford},\ \bibinfo {year} {2000})\BibitemShut {NoStop}%
\bibitem [{\citenamefont {Agarwal}(2012)}]{Agarwal_2012}%
  \BibitemOpen
  \bibfield  {author} {\bibinfo {author} {\bibfnamefont {G.~S.}\ \bibnamefont {Agarwal}},\ }\href {https://doi.org/10.1017/cbo9781139035170} {\emph {\bibinfo {title} {Quantum Optics}}}\ (\bibinfo  {publisher} {Cambridge University Press},\ \bibinfo {address} {Cambridge, England},\ \bibinfo {year} {2012})\BibitemShut {NoStop}%
\bibitem [{\citenamefont {Demkowicz-Dobrza{\'n}ski}\ \emph {et~al.}(2015)\citenamefont {Demkowicz-Dobrza{\'n}ski}, \citenamefont {Jarzyna},\ and\ \citenamefont {Ko{\l}ody{\'n}ski}}]{demkowicz2015quantum}%
  \BibitemOpen
  \bibfield  {author} {\bibinfo {author} {\bibfnamefont {R.}~\bibnamefont {Demkowicz-Dobrza{\'n}ski}}, \bibinfo {author} {\bibfnamefont {M.}~\bibnamefont {Jarzyna}},\ and\ \bibinfo {author} {\bibfnamefont {J.}~\bibnamefont {Ko{\l}ody{\'n}ski}},\ }\bibfield  {title} {\bibinfo {title} {Quantum limits in optical interferometry},\ }\href@noop {} {\bibfield  {journal} {\bibinfo  {journal} {Progress in Optics}\ }\textbf {\bibinfo {volume} {60}},\ \bibinfo {pages} {345} (\bibinfo {year} {2015})}\BibitemShut {NoStop}%
\bibitem [{\citenamefont {Paris}(2009)}]{paris2009quantum}%
  \BibitemOpen
  \bibfield  {author} {\bibinfo {author} {\bibfnamefont {M.~G.}\ \bibnamefont {Paris}},\ }\bibfield  {title} {\bibinfo {title} {Quantum estimation for quantum technology},\ }\href@noop {} {\bibfield  {journal} {\bibinfo  {journal} {International Journal of Quantum Information}\ }\textbf {\bibinfo {volume} {7}},\ \bibinfo {pages} {125} (\bibinfo {year} {2009})}\BibitemShut {NoStop}%
\bibitem [{\citenamefont {Royer}(1977)}]{royer1977wigner}%
  \BibitemOpen
  \bibfield  {author} {\bibinfo {author} {\bibfnamefont {A.}~\bibnamefont {Royer}},\ }\bibfield  {title} {\bibinfo {title} {Wigner function as the expectation value of a parity operator},\ }\href@noop {} {\bibfield  {journal} {\bibinfo  {journal} {Physical Review A}\ }\textbf {\bibinfo {volume} {15}},\ \bibinfo {pages} {449} (\bibinfo {year} {1977})}\BibitemShut {NoStop}%
\bibitem [{\citenamefont {Cahill}\ and\ \citenamefont {Glauber}(1969)}]{cahill1969density}%
  \BibitemOpen
  \bibfield  {author} {\bibinfo {author} {\bibfnamefont {K.~E.}\ \bibnamefont {Cahill}}\ and\ \bibinfo {author} {\bibfnamefont {R.~J.}\ \bibnamefont {Glauber}},\ }\bibfield  {title} {\bibinfo {title} {Density operators and quasiprobability distributions},\ }\href@noop {} {\bibfield  {journal} {\bibinfo  {journal} {Physical Review}\ }\textbf {\bibinfo {volume} {177}},\ \bibinfo {pages} {1882} (\bibinfo {year} {1969})}\BibitemShut {NoStop}%
\bibitem [{\citenamefont {Gerry}\ \emph {et~al.}(2007)\citenamefont {Gerry}, \citenamefont {Benmoussa},\ and\ \citenamefont {Campos}}]{gerry2007parity}%
  \BibitemOpen
  \bibfield  {author} {\bibinfo {author} {\bibfnamefont {C.~C.}\ \bibnamefont {Gerry}}, \bibinfo {author} {\bibfnamefont {A.}~\bibnamefont {Benmoussa}},\ and\ \bibinfo {author} {\bibfnamefont {R.}~\bibnamefont {Campos}},\ }\bibfield  {title} {\bibinfo {title} {Parity measurements, heisenberg-limited phase estimation, and beyond},\ }\href@noop {} {\bibfield  {journal} {\bibinfo  {journal} {Journal of Modern Optics}\ }\textbf {\bibinfo {volume} {54}},\ \bibinfo {pages} {2177} (\bibinfo {year} {2007})}\BibitemShut {NoStop}%
\bibitem [{\citenamefont {Bollinger}\ \emph {et~al.}(1996)\citenamefont {Bollinger}, \citenamefont {Itano}, \citenamefont {Wineland},\ and\ \citenamefont {Heinzen}}]{PhysRevA.54.R4649}%
  \BibitemOpen
  \bibfield  {author} {\bibinfo {author} {\bibfnamefont {J.~J.~.}\ \bibnamefont {Bollinger}}, \bibinfo {author} {\bibfnamefont {W.~M.}\ \bibnamefont {Itano}}, \bibinfo {author} {\bibfnamefont {D.~J.}\ \bibnamefont {Wineland}},\ and\ \bibinfo {author} {\bibfnamefont {D.~J.}\ \bibnamefont {Heinzen}},\ }\bibfield  {title} {\bibinfo {title} {Optimal frequency measurements with maximally correlated states},\ }\href {https://doi.org/10.1103/PhysRevA.54.R4649} {\bibfield  {journal} {\bibinfo  {journal} {Phys. Rev. A}\ }\textbf {\bibinfo {volume} {54}},\ \bibinfo {pages} {R4649} (\bibinfo {year} {1996})}\BibitemShut {NoStop}%
\bibitem [{\citenamefont {Feng}\ \emph {et~al.}(2014)\citenamefont {Feng}, \citenamefont {Jin},\ and\ \citenamefont {Yang}}]{feng2014quantum}%
  \BibitemOpen
  \bibfield  {author} {\bibinfo {author} {\bibfnamefont {X.}~\bibnamefont {Feng}}, \bibinfo {author} {\bibfnamefont {G.}~\bibnamefont {Jin}},\ and\ \bibinfo {author} {\bibfnamefont {W.}~\bibnamefont {Yang}},\ }\bibfield  {title} {\bibinfo {title} {Quantum interferometry with binary-outcome measurements in the presence of phase diffusion},\ }\href@noop {} {\bibfield  {journal} {\bibinfo  {journal} {Physical Review A}\ }\textbf {\bibinfo {volume} {90}},\ \bibinfo {pages} {013807} (\bibinfo {year} {2014})}\BibitemShut {NoStop}%
\bibitem [{\citenamefont {Cohen}\ \emph {et~al.}(2014)\citenamefont {Cohen}, \citenamefont {Istrati}, \citenamefont {Dovrat},\ and\ \citenamefont {Eisenberg}}]{cohen2014super}%
  \BibitemOpen
  \bibfield  {author} {\bibinfo {author} {\bibfnamefont {L.}~\bibnamefont {Cohen}}, \bibinfo {author} {\bibfnamefont {D.}~\bibnamefont {Istrati}}, \bibinfo {author} {\bibfnamefont {L.}~\bibnamefont {Dovrat}},\ and\ \bibinfo {author} {\bibfnamefont {H.}~\bibnamefont {Eisenberg}},\ }\bibfield  {title} {\bibinfo {title} {Super-resolved phase measurements at the shot noise limit by parity measurement},\ }\href@noop {} {\bibfield  {journal} {\bibinfo  {journal} {Optics express}\ }\textbf {\bibinfo {volume} {22}},\ \bibinfo {pages} {11945} (\bibinfo {year} {2014})}\BibitemShut {NoStop}%
\bibitem [{\citenamefont {Cohen}\ \emph {et~al.}(2013)\citenamefont {Cohen}, \citenamefont {Istrati}, \citenamefont {Dovrat},\ and\ \citenamefont {Eisenberg}}]{cohen2013experimental}%
  \BibitemOpen
  \bibfield  {author} {\bibinfo {author} {\bibfnamefont {L.}~\bibnamefont {Cohen}}, \bibinfo {author} {\bibfnamefont {D.}~\bibnamefont {Istrati}}, \bibinfo {author} {\bibfnamefont {L.}~\bibnamefont {Dovrat}},\ and\ \bibinfo {author} {\bibfnamefont {H.}~\bibnamefont {Eisenberg}},\ }\bibfield  {title} {\bibinfo {title} {Experimental super-resolved phase measurements with shot-noise sensitivity},\ }\href@noop {} {\bibfield  {journal} {\bibinfo  {journal} {arXiv preprint arXiv:1311.2721}\ } (\bibinfo {year} {2013})}\BibitemShut {NoStop}%
\bibitem [{\citenamefont {Wang}\ \emph {et~al.}(2016{\natexlab{b}})\citenamefont {Wang}, \citenamefont {Hao}, \citenamefont {Zhang}, \citenamefont {Xu}, \citenamefont {Yang}, \citenamefont {Yang},\ and\ \citenamefont {Zhao}}]{wang2016super}%
  \BibitemOpen
  \bibfield  {author} {\bibinfo {author} {\bibfnamefont {Q.}~\bibnamefont {Wang}}, \bibinfo {author} {\bibfnamefont {L.}~\bibnamefont {Hao}}, \bibinfo {author} {\bibfnamefont {Y.}~\bibnamefont {Zhang}}, \bibinfo {author} {\bibfnamefont {L.}~\bibnamefont {Xu}}, \bibinfo {author} {\bibfnamefont {C.}~\bibnamefont {Yang}}, \bibinfo {author} {\bibfnamefont {X.}~\bibnamefont {Yang}},\ and\ \bibinfo {author} {\bibfnamefont {Y.}~\bibnamefont {Zhao}},\ }\bibfield  {title} {\bibinfo {title} {Super-resolving quantum lidar: entangled coherent-state sources with binary-outcome photon counting measurement suffice to beat the shot-noise limit},\ }\href@noop {} {\bibfield  {journal} {\bibinfo  {journal} {Optics express}\ }\textbf {\bibinfo {volume} {24}},\ \bibinfo {pages} {5045} (\bibinfo {year} {2016}{\natexlab{b}})}\BibitemShut {NoStop}%
\bibitem [{\citenamefont {Dowling}\ and\ \citenamefont {Seshadreesan}(2014)}]{dowling2014quantum}%
  \BibitemOpen
  \bibfield  {author} {\bibinfo {author} {\bibfnamefont {J.~P.}\ \bibnamefont {Dowling}}\ and\ \bibinfo {author} {\bibfnamefont {K.~P.}\ \bibnamefont {Seshadreesan}},\ }\bibfield  {title} {\bibinfo {title} {Quantum optical technologies for metrology, sensing, and imaging},\ }\href@noop {} {\bibfield  {journal} {\bibinfo  {journal} {Journal of Lightwave Technology}\ }\textbf {\bibinfo {volume} {33}},\ \bibinfo {pages} {2359} (\bibinfo {year} {2014})}\BibitemShut {NoStop}%
\bibitem [{\citenamefont {Malik}\ and\ \citenamefont {Boyd}(2014)}]{malik2014quantum}%
  \BibitemOpen
  \bibfield  {author} {\bibinfo {author} {\bibfnamefont {M.}~\bibnamefont {Malik}}\ and\ \bibinfo {author} {\bibfnamefont {R.~W.}\ \bibnamefont {Boyd}},\ }\bibfield  {title} {\bibinfo {title} {Quantum imaging technologies},\ }\href@noop {} {\bibfield  {journal} {\bibinfo  {journal} {La Rivista del Nuovo Cimento}\ }\textbf {\bibinfo {volume} {37}},\ \bibinfo {pages} {273} (\bibinfo {year} {2014})}\BibitemShut {NoStop}%
\bibitem [{\citenamefont {Qiang}\ \emph {et~al.}(2018)\citenamefont {Qiang}, \citenamefont {Lili}, \citenamefont {Hongxia}, \citenamefont {Xianli}, \citenamefont {Haiwei}, \citenamefont {Lianfu},\ and\ \citenamefont {Yuan}}]{qiang2018effects}%
  \BibitemOpen
  \bibfield  {author} {\bibinfo {author} {\bibfnamefont {W.}~\bibnamefont {Qiang}}, \bibinfo {author} {\bibfnamefont {H.}~\bibnamefont {Lili}}, \bibinfo {author} {\bibfnamefont {T.}~\bibnamefont {Hongxia}}, \bibinfo {author} {\bibfnamefont {L.}~\bibnamefont {Xianli}}, \bibinfo {author} {\bibfnamefont {M.}~\bibnamefont {Haiwei}}, \bibinfo {author} {\bibfnamefont {H.}~\bibnamefont {Lianfu}},\ and\ \bibinfo {author} {\bibfnamefont {Z.}~\bibnamefont {Yuan}},\ }\bibfield  {title} {\bibinfo {title} {Effects of real environments on the performance of quantum lidar},\ }\href@noop {} {\bibfield  {journal} {\bibinfo  {journal} {Infrared Laser Eng.}\ }\textbf {\bibinfo {volume} {47}},\ \bibinfo {pages} {29} (\bibinfo {year} {2018})}\BibitemShut {NoStop}%
\end{thebibliography}%

\end{document}